\documentclass[11pt]{article}
\usepackage{amsmath}
\usepackage{epsfig}
\usepackage{graphicx}
\newtheorem{Lemma}{Lemma}
\newtheorem{Theorem}[Lemma]{Theorem}
\newtheorem{Proposition}[Lemma]{Proposition}
\newcommand{\mrca}{\mbox{\rm MRCA}}
\newcommand{\TREE}{\mathrm{TREE}}
\newcommand{\GENERA}{\mathrm{GENERA}}
\newcommand{\GENUS}{\mathrm{GENUS}}
\newcommand{\type}{\mathrm{type}}
\newcommand{\Torigin}{T^{\mbox{\small{origin}}}}
\newcommand{\Tmrca}{T^{\mbox{\small{mrca}}}}
\newcommand{\TT}{\mbox{${\cal T}$}}
\newcommand{\LL}{\mbox{${\cal L}$}}
\newcommand{\sqt}{\sqrt\theta}
\newcommand{\sfrac}[2]{{\textstyle\frac{#1}{#2}}}
\newcommand{\GG}{{\mathcal G}}
\newcommand{\Th}{\theta}

\def\bA{{\bar A}}
\def\bB{{\bar B}}

\begin{document}
\title{Stochastic Models for Phylogenetic Trees on Higher-order Taxa}
\author{David Aldous\thanks{Research supported by NSF Grant DMS-0704159} 
\\
Department of Statistics\\
367 Evans Hall \# 3860\\
U.C. Berkeley, CA 94720-3860\\
aldous@stat.berkeley.edu
\and 
Maxim Krikun \\
Institut Elie Cartan\\
Universite Henri Poincare\\ 
Nancy, France \\
krikun@iecn.u-nancy.fr 
\and 
Lea Popovic\\
Dept of Mathematics \\
Cornell University \\
Ithaca NY 14853, USA \\
e-mail: popovic@math.cornell.edu 
}
\maketitle

\begin{abstract}
Simple stochastic models
for phylogenetic trees on species 
have been well studied.
But much paleontology data concerns time series or trees on higher-order taxa, 
and any broad picture of relationships between extant groups 
requires use of higher-order taxa.
A coherent model for trees on (say) genera 
should involve
both a species-level model and a model for the classification scheme 
by which species are assigned to genera.
We present a general framework for such models, and describe three alternate classification schemes.
Combining with the species-level model of Aldous-Popovic (2005),
one gets models for higher-order trees, and we initiate analytic study of such models.
In particular we derive formulas for the lifetime of genera, for the distribution of number of species per genus,
and for the offspring structure of the tree on genera.
\end{abstract}

\newpage
\section{Introduction}
This paper provides some mathematical details of part of a broader project
we call ``coherent modeling of macroevolution". 
The focus here is on a novel mathematical framework and on analytical results for a model of macroevolution which is possible within this framework.
We give only a very brief sketch of the motivation for the project in section \ref{sec-overview} and then proceed to outline (section \ref{sec-results})
the specific results of this paper.
A future paper \cite{me-macro}
addressed to less mathematically-focused biologists
will provide more detailed background, motivation, relation with previous biological 
literature, and discuss the ``bottom line for biology" of such mathematical models.

\subsection{Background}
\label{sec-overview}
Stochastic models for time series of species numbers within a clade
and for phylogenetic trees of extant species in a clade
can be traced back to Yule \cite{yule24}.
Such models treat speciations and extinctions as random (in some way).
In studying such models one is not asserting that real macroevolution was
purely random; rather, one wishes to compare real data with the predictions
of a random model to see what patterns require biological explanation 
(e.g. adaptive radiations \cite{Guyer93}),
or to make inference about unobservables 
(e.g. the time of origin of the primates \cite{TMW02}).

One aspect of this subject is where the data consists of time series or phylogenetic
trees on some higher-level taxa (genera or families, say) instead of species.
In the fossil record of the distant past it is 
difficult to resolve specimens to the species level, and the species-level data is liable to
be incomplete, so that statistical analysis
of time series
(relying e.g. on the celebrated compendia of Sepkoski \cite{sepkoski92})
is in practice done at the level of genera or families.
In discussing phylogenies within large extant groups such as birds or mammals, 
it is impractical to show all species, so one shows trees indicating how
major subgroups are related.
And the same holds for extinct groups
(see e.g. the fascinating tree \cite{PYLB02}
on dinosaur genera).
In looking at such data and asking the basic question -- what patterns imply
biological effect rather than being consistent with ``just chance" -- 
two extra difficulties arise.  
First, the classification into genera or families involves human judgement
which is inevitably at least somewhat subjective.
Second, while one could just take genera (say) as entities in themselves and
apply species-level models directly to genera \cite{gouldSE},
this seems conceptually unsatisfactory:
genera are sets of species and so, 
as part of ``{\em coherent} modeling of macroevolution",
one would like genus-level models to be based upon underlying species-level models.
While these difficulties are certainly mentioned in the biological literature,
we have not seen any very thorough mathematical study.
Our purposes in this paper are to lay out a conceptual framework for studying
such questions, and then to give mathematical analysis of the predictions of
a particular probability model.

\subsection{The two topics of this paper}
\label{sec-results}
The first topic concerns methodology of classification 
(without any involvement of probability models).
Suppose we know the true phylogenetic tree on a {\em clade} of species, that is
on some founder species $s$ (typically extinct) and the set of
all descendant species (extant and extinct) of $s$.  
How might one assign species to genera?
(From now on we write {\em genus, genera} for concreteness to indicate any 
higher level of the taxonomic hierarchy).
Suppose we distinguish certain species as ``new type"
due to some characteristic judged biologically significant
which persists in descendant species. 
Then it seems sensible to use these as a basis for classification -- very
roughly, a ``new type" species is the founder of a new genus.
This set-up ignores various practical problems 
(one seldom has the complete tree on extinct species; 
which characteristics should one choose as significant?)
but does lead us to a purely mathematical question:
\begin{quote} 
Consider a {\em classification scheme} which, given any phylogenetic tree in which some
species are distinguished as ``new type", classifies all species into genera.
What desirable properties can such schemes have?
\end{quote}
Section \ref{sec-class} gives our answer.
One might hope there was some single mathematically natural scheme, but it turns
out that different desiderata dictate different schemes.  
We pick out three schemes which we name {\em fine, medium, coarse}
and describe their properties.

Studying to what extent actual taxonomic practice resembles one of these theoretical schemes
would make an interesting project in statistical analysis, but this is not our 
purpose here.  Instead, as our second topic we use these theoretical classification schemes to consider the questions:
\begin{itemize} \vspace{-3mm}
\item 
In what ways might phylogenetic trees on genera, or time series of genera,
differ from those on species? \vspace{-3mm}
\item In what ways might phylogenetic trees on species in the same $n$-species genus differ from phylogenetic trees on species of simply an $n$-species clade?  
(Section \ref{sec-pmhot} amplifies the issue here.) \vspace{-3mm}
\item  How does the choice of classification scheme for 
determining genera affect such differences? \vspace{-3mm}
\end{itemize}
We study these questions under a certain probability model
for the underlying phylogenetic tree on species.
This model, described in section \ref{sec-species-level} and studied in
detail in \cite{me111},
is intended to formalize the idea of 
``purely random" history subject to a given number $n$ of extant species.
Section \ref{sec-analysis} investigates the statistical structure of phylogenies on genera
obtained by combining the species-level model with the
genera classification schemes, and this combination is the conceptual novelty of the paper.
In particular we derive formulas (Theorem \ref{T2}) for the lifetime of genera and for the distribution of number of species per genus,
and formulas (Propositions \ref{P4} and \ref{P5}) 
for the offspring structure of the tree on genera,
both of these results in the (mathematically easier) case of extinct clades;
and (for extant clades) the number of species per genus (Proposition \ref{P6}).
As noted in section \ref{sec-analysis} there are many more calculations one might attempt to perform, 
and we invite interested readers to extend our calculations.

\section{Phylogenetic trees and genera classification schemes}
\label{sec-class}
\subsection{Cladograms}
\label{sec-clad}
For good reasons, both practical and theoretical, 
phylogenetic relationships are usually presented via a {\em cladogram},
a binary tree (cf. Figure 4) without time scale and without identifying branchpoints with
explicit taxa.
A mathematical discussion of genera classification schemes
would be simpler if it were based only on the reduced information provided by cladograms on species.
But our goal is to see how 
phylogenies on genera emerge from some complete underlying process of macroevolution at the species level in which species originate and go extinct at particular times.
This requires using a species-level model on {\em phylogenetic trees} as defined below,
even though ultimately one may 
choose to express relationships between genera using cladograms.

\subsection{Phylogenetic trees}
\label{sec-pt}
Our basic assumptions about macroevolution in a clade of species are logically simple,
although oversimplified in reality. 
\begin{itemize} \vspace{-3mm}
\item Each species has a ``time of origin" and either is extant or has a ``time of extinction"; \vspace{-3mm}
\item Each species (except the founder of the clade) originates as a ``daughter" of 
some ``parent" species in the clade, not simultaneously with any other daughter.
\end{itemize} \vspace{-3mm}

\newpage
\setlength{\unitlength}{0.1in}
\begin{picture}(50,35)
\put(0,12){\line(0,1){22}}
\put(-0.4,11.03){a}
\put(1,4){\line(0,1){10}}
\put(0.6,3.03){b}
\put(2,0){\line(0,1){6}}
\put(1.6,-0.97){c}
\put(3,0){\line(0,1){8}}
\put(2.6,-0.97){d}
\put(4,10){\line(0,1){9}}
\put(3.6,9.03){e}
\put(5,13){\line(0,1){2}}
\put(4.6,12.03){f}
\put(6,7){\line(0,1){10}}
\put(5.6,6.03){g}
\put(7,3){\line(0,1){6}}
\put(6.6,2.03){h}
\put(8,5){\line(0,1){6}}
\put(7.6,4.03){i}
\put(9,15){\line(0,1){1}}
\put(8.6,14.03){j}
\put(10,21){\line(0,1){1}}
\put(9.6,20.03){k}
\put(11,26){\line(0,1){6}}
\put(10.8,24.5){l}
\put(12,18){\line(0,1){11}}
\put(11.6,17.03){m}
\put(13,19){\line(0,1){1}}
\put(12.6,18.03){n}
\put(14,8){\line(0,1){16}}
\put(13.6,7.03){o}
\put(15,4){\line(0,1){8}}
\put(14.6,3.03){p}
\put(16,0){\line(0,1){7}}
\put(15.3,-0.97){q}
\put(17,0){\line(0,1){10}}
\put(16.6,-0.97){r}
\put(18,9){\line(0,1){8}}
\put(17.6,8.03){s}
\put(19,0){\line(0,1){13}}
\put(18.6,-0.97){t}
\put(20,5){\line(0,1){10}}
\put(19.6,4.03){u}
\put(21,0){\line(0,1){8}}
\put(20.6,-0.97){v}
\put(1,14){\line(-1,0){1}}
\put(2,6){\line(-1,0){1}}
\put(3,8){\line(-1,0){2}}
\put(4,19){\line(-1,0){4}}
\put(5,15){\line(-1,0){1}}
\put(6,17){\line(-1,0){2}}
\put(7,9){\line(-1,0){1}}
\put(8,11){\line(-1,0){2}}
\put(7.5,10.6){$\bullet$}
\put(9,16){\line(-1,0){3}}
\put(10,22){\line(-1,0){10}}
\put(11,32){\line(-1,0){11}}
\put(12,29){\line(-1,0){1}}
\put(13,20){\line(-1,0){1}}
\put(14,24){\line(-1,0){2}}
\put(15,12){\line(-1,0){1}}
\put(16,7){\line(-1,0){1}}
\put(17,10){\line(-1,0){2}}
\put(18,17){\line(-1,0){4}}
\put(17.5,16.6){$\bullet$}
\put(19,13){\line(-1,0){1}}
\put(20,15){\line(-1,0){2}}
\put(21,8){\line(-1,0){1}}
\put(30,0){\line(0,1){34}}
\put(28.1,-1){abcdk}
\put(33,10){\line(0,1){9}}
\put(32.4,9){ef}
\put(36,3){\line(0,1){14}}
\put(35,1.8){ghj}
\put(39,5){\line(0,1){6}}
\put(38.6,4){i}
\put(42,26){\line(0,1){6}}
\put(41.8,24.5){l}
\put(45,18){\line(0,1){11}}
\put(44,17){mn}
\put(48,0){\line(0,1){24}}
\put(46.3,-1){opqr}
\put(51,0){\line(0,1){17}}
\put(50,-1){stuv}
\put(33,19){\line(-1,0){3}}
\put(36,17){\line(-1,0){3}}
\put(39,11){\line(-1,0){3}}
\put(42,32){\line(-1,0){12}}
\put(45,29){\line(-1,0){3}}
\put(48,24){\line(-1,0){3}}
\put(51,17){\line(-1,0){3}}
\put(8,36){{\bf species}}
\put(37,36){{\bf fine genera}}
\end{picture}

\vspace{0.2in}

{\bf Figure 1.}
{\small 
Illustration of our schemes for defining genera in terms of
new types.
Above left is a complete clade of 6 extant and 16 extinct species
$(abcd \cdots uv)$,
with two species $\{i,s\}$ designated as
new types
and marked $\bullet$.
In the fine scheme,
this induces 8 genera (3 extant), whose tree is shown above right.
The other schemes are shown below, with compressed time scale.}\\

\setlength{\unitlength}{0.1in}
\begin{picture}(50,21)
\put(30,0){\line(0,1){17}}
\put(26.1,-1){abcdefghjk}
\put(36,2.5){\line(0,1){3}}
\put(35.6,1.5){i}
\put(42,0){\line(0,1){14.5}}
\put(38,-1){lmnopqr}
\put(46,0){\line(0,1){8.5}}
\put(45,-1){stuv}
\put(36,5.5){\line(-1,0){6}}
\put(42,14.5){\line(-1,0){12}}
\put(46,8.5){\line(-1,0){4}}
\put(37,17){{\bf medium genera}}
\put(0,0){\line(0,1){17}}
\put(6,2.5){\line(0,1){3}}
\put(5.6,1.5){i}
\put(12,0){\line(0,1){8.5}}
\put(11,-1){stuv}
\put(6,5.5){\line(-1,0){6}}
\put(12,8.5){\line(-1,0){12}}
\put(4,17){{\bf coarse genera}}
\put(-2,-1){other}
\end{picture}

\newpage
A {\em phylogenetic tree} records all this information
(birth and extinction times; parent-daughter relationships).
There are many different ways to draw such a tree.
Figure 1 (top left) uses one convention, explained further in Figure 2:  time increases downwards,
a species is indicated by a vertical line from time of origin to time of extinction 
or the present time, and parent-daughter relationship is indicated by a horizontal line
with the daughter on the right.

For later use in proofs we state some language for discussing this particular
representation of phylogenetic trees.
Given a parent-daughter pair, there is a 
{\em branchpoint} on the parent's line, from which a
{\em right edge} leads to the daughter and a
{\em continuing edge} leads down to another branchpoint or 
the {\em leaf} representing extinction time of
the parent species or the current time.
See Figure 2.

\setlength{\unitlength}{0.37in}
\begin{picture}(8,6.4)(-2.5,0)
\put(0,1){\line(0,1){4}}
\put(0,4){\line(1,0){6}}
\put(2,0){\line(0,1){1.5}}
\put(0,1.5){\line(1,0){2}}
\put(6,0){\line(0,1){4}}
\put(6,2){\line(1,0){2}}
\put(8,2){\line(0,-1){1.5}}
\put(-0.3,5.2){{\small parent}}
\put(5.4,4.2){{\small daughter}}
\put(2.0,4.2){{\small right edge}}
\put(-2.1,3.95){{\small branchpoint}}
\put(-1.0,2.98){{\small continuing}}
\put(-0.5,2.7){{\small edge}}
\put(-0.7,0.9){{\small leaf}}
\end{picture}

\vspace{0.2in}

{\bf Figure 2.}
{\small 
Terminology for edges of phylogenetic trees.}

\vspace{0.12in}
\noindent
Any species determines a {\em subclade} consisting of itself
and all its descendant species.
Similarly a continuing edge determines a subclade 
(consisting of the species whose line contains the continuing edge, 
and later daughter species and their descendants).

\subsection{Discussion} 
Of course the ``basic assumptions" above represent one extreme of the various mechanisms of speciation discussed by biologists -- that speciation typically arises from innovation, in such a way that there is a new lineage splitting off from an
old lineage which continues unchanged.  
This type of lineage splitting is generally considered plausible for
fast-evolving organisms such as viruses and bacteria, but its plausibility for
macrofauna is more debatable.  We chose this both for definiteness and because it seems most amenable to mathematical modelling.  

This paper is framed in the context of 
phylogenetic tree structure assigned to traditional rank-based taxa.  
But one could alternatively frame it  within the proposed Phylocode \cite{phylocode} conventions for 
naming the parts of the tree of life by explicit reference to phylogeny.  
While Phylocode provides a logical representation of the fine detail of relationships between all species, 
the  high-level structure 
-- what we want to teach, starting in grade school with the relationship between mammals, birds and fish -- 
requires us in practice to distinguish some clades as important and then to exhibit trees showing their relationship.
So in the sequel one can interpret  ``genus" as  ``clade distinguished as important for the purposes of 
exhibiting high-level structure of the tree of life".

\subsection{Desirable properties for genera classification schemes}
\label{sec-DP}
As mentioned earlier, the question we study in section \ref{sec-class}
is:
\begin{quote}
given a phylogenetic tree and a subset of its species designated as ``new type", 
how can one classify all species into genera?
\end{quote}
We start by considering three desirable properties for classification schemes.
For all our schemes we require the following weak
formalization of the idea that ``new type" species should initiate new genera.

\noindent
{\bf Property 1.}
A genus cannot contain both a species $a$ which is a descendant of some 
``new type" species $s$ and also a species $b$ which is not a descendant of $s$.

Here ``descendant" includes $s$ itself, so in particular a 
``new type" species and its parent must be in different genera.

Next note that if we required every genus to be a clade
(monophyletic) then we could never have more than one genus, because
otherwise some parent-daughter pair $\{a,b\}$ would be in different genera 
and then the genus containing $a$ is not a clade.
We will consider a weaker property.
Any two distinct species $a,b$ have a
{\em most recent common ancestor} 
$\mrca(a,b)$,
which is some species (maybe $a$ or $b$).
Given three distinct species $\{a,b,c\}$, say
{\em $(a,b)$ are more closely related than $(a,c)$}
if $\mrca(a,b)$ is a descendant of $\mrca(a,c)$.
Here again we allow
$\mrca(a,b) = \mrca(a,c)$.


\noindent
{\bf Property 2.}
Given three distinct species $\{a,b,c\}$, 
with $a$ and $b$ in the same genus and $c$ in a different genus,
then
$(a,b)$ are more closely related than $(a,c)$.

As another kind of desirable property,
one would like to be able to draw a tree or cladogram on 
{\em genera} is some unique way, and the next property (for a
classification scheme) provides one formalization of this idea.

\noindent
{\bf Property 3.}
Choosing one representative species from each genus and drawing the cladogram on these
species gives a cladogram which does not depend on choice of representative species.

\subsection{Three genera classification schemes}
\label{sec-3g}
The properties above are always satisfied by the trivial scheme in
which each species is declared to be a separate genus.
Roughly speaking, such properties become easier to satisfy if one
uses more genera, 
so one should consider
schemes which produce the {\em smallest} number of genera
consistent with specified properties.
More precisely, if 
$(G_i)$ and $(G^\prime_j)$ are two different classifications of
the same set of species into genera, say
{\em $(G^\prime_j)$ is coarser than $(G_i)$}
if each $G^\prime_j$ is the union of one or more of the $G_i$.
Our main result in section \ref{sec-class}, Theorem \ref{T1},
gives explicit constructions of the coarsest genera classification
schemes satisfying various properties.

Observe that any way of attaching ``marks" to some {\em edges}
of the phylogenetic tree
(in the representation of section \ref{sec-pt})
can be used to define genera, by declaring that species
$\{a,b\}$ are in the same genus if and only if the path
in the tree from the leaf $a$ to the leaf $b$ contains 
no marked edge.
Here are three ways one might attach marks to edges.

(a) At each parent-daughter branchpoint where the daughter is
``new type",
mark the right edge (from parent to daughter).

(b) At each parent-daughter branchpoint where the daughter's
subclade contains some
``new type" species,
mark the right edge (from parent to daughter).

(c) At each parent-daughter branchpoint where the daughter's
subclade and the continuing edge subclade both contain some
``new type" species,
mark both the right edge and the continuing edge.

Now define three genera classification schemes as follows.

\noindent
{\bf Coarse scheme:} create marks according to rule (a).

\noindent
{\bf Medium scheme:} create marks according to rules (a) and (c).

\noindent
{\bf Fine scheme:} create marks according to rules (a) and (b).

In each case the marks define genera as above.
Figure 9 provides a visual catalog of these rules. 
\begin{Theorem}
\label{T1}
(i) The {\bf coarse} scheme defines genera with 
Property 1,
and is coarser than any other scheme satisfying
Property 1.\\
(ii) The {\bf fine} scheme defines genera with 
Properties 1 and 2,
and is coarser than any other scheme satisfying
Properties 1 and 2.\\ 
(iii) The {\bf medium} scheme defines genera with 
Properties 1 and 3,
and is coarser than any other scheme satisfying
Properties 1 and 3. 
\end{Theorem}
This is proved in the next section.
Section  \ref{sec-appendix} contains further discussion, in particular on the 
paraphyletic property of these genera.

\paragraph{Remarks}
 Recall that, in these classification schemes, the data we start with is a
phylogenetic tree with certain species distinguished as ``new type".  The marks
described above are artifacts used in the algorithmic construction of genera and
in their analysis.  In particular, in Figure 2 the continuing edge (representing
a continuation of the same species) cannot represent a new type species, even
though (in the medium scheme) it may be given a mark.   Unlike in cladograms,
the two edges following a branchpoint in the underlying phylogenetic tree are
not interchangeable.

One may well object that these classification schemes do not
correspond to the ways in which systematists actually assign
taxonomic ranks; but we do not know any discussion of the latter
in the biological literature which is amenable to mathematical
modeling.  Recall that our ultimate goal is to compare real
evolutionary history to the predictions of some ``pure chance"
model to see what differences can be found.   Having several
alternate choices for the genera classification part of the model seems helpful, in
that a difference  in consistent direction from all the models seems more worthy of
consideration.

\subsection{Proof of Theorem \ref{T1}}
Showing that the schemes define genera with the stated properties
is straightforward, as follows.

{\em Case (i)}.
The marks from rule (a) ensure Property 1.

{\em Case (ii)}.
Consider genera defined using marks from rules (a) and (b).
Suppose $\{a,b\}$ are in the same genus and $c$ is in a different genus.
We'll prove by contradiction that $(a,b)$ are necessarily more closely
related than $(a,c)$.
If $(a,b)$ are not more closely related than $(a,c)$,
there exist a species $s$ such that both $a$ and $c$ are in the 
subclade of $s$, while $b$ is not (it's possible that $a=s$ or $c=s$).
But the path from $a$ to $c$ contains a marked edge, meaning that there
is at least one new type species in the subclade of $s$.
According to rule (b), the edge between $s$ and its parent $s'$ is marked,
and because $b$ is not in the subclade of $s$, it cannot be in the same 
genus as $a$.

{\em Case (iii)}.
Let $\{a,b,c,\ldots\}$ 
be representatives of the different genera,
and let $a^\prime$ be in the same genus as $a$. 
We need to show that the cladograms on
$\{a,b,c,\ldots\}$ 
and on
$\{a^\prime,b,c,\ldots\}$ 
are the same. 
Consider the cladogram on
$\{a,a^\prime,b,c,\ldots\}$.
In this cladogram consider the branchpoint above the leaf $a$ 
and the branchpoint above the leaf $a^\prime$.
If these branchpoints are the same or or linked by an edge then the two cladograms
are indeed identical. 
Otherwise we have a cladogram as in Figure 3.

\setlength{\unitlength}{0.28in}
\begin{picture}(8,4.2)(-5,0)
\put(0,0){\line(1,1){4}}
\put(2,0){\line(-1,1){1}}
\put(4,0){\line(-1,1){2}}
\put(6,0){\line(-1,1){3}}
\put(-0.1,-0.5){a}
\put(1.8,-0.5){b}
\put(3.8,-0.5){c}
\put(5.8,-0.5){$a^\prime$}
\put(1.3,1.7){$\times$}
\put(1.3,0.4){$\times$}
\put(2.3,1.4){$\times$}
\end{picture}

\vspace{0.2in}

{\bf Figure 3.}
{\small 
Cladogram arising in the proof of case (iii).
}

\vspace{0.1in}
\noindent
Edges in the cladogram can be identified with paths in the phylogenetic tree.
Because $a$ and $a^\prime$ are in the same genus,
there is no mark on the path from $a$ to $a^\prime$.
Because $b$ and $c$ are in different genera, there is a mark on the
cladogram edges to $b$ and to $c$.
But then by examining rules (a) and (c) we see there must be a mark on
the cladogram edge from 
$\mrca(a,b)$ to $\mrca(a,c)$,
contradicting the assumption that $a$ and $a^\prime$ are in the same genus.
This verifies case (iii).

We now need to prove the ``coarser than" assertions.
In each case, it is enough to show that
if $G$ is a genus in some scheme satisfying the relevant properties,
then it is part of a genus in the specified scheme 
(coarse, medium, fine).
In other words, we need to show that if
$a$ and $b$ are in the same 
genus in some scheme satisfying the relevant properties,
then the path from the leaf of $a$ to the leaf of $b$ 
in the phylogenetic tree does not contain any marks of the relevant kind.
We will argue by contradiction, supposing that some edge
$(c,d)$ on the path {\em does} have a mark.

{\em Case (i)}.
Here $(c,d)$ is a parent-daughter edge and $d$ is a
``new type".
One of $\{a,b\}$ -- say $b$ -- is in the subclade of $d$,
and so $a$ is not in that subclade.  
But this violates Property 1.

{\em Case (ii)}.
Again $(c,d)$ is a parent-daughter edge, 
and we may assume $b$ is in the subclade of $d$,
and so $a$ is not in that subclade.  
Also some species $f$ in the subclade of $d$ is a ``new type" species.
By Property 1 $f$ is in a different genus from $a$,
then by Property 2
$\mrca(a,b)$ is a descendant of $\mrca(b,f)$.
But this is impossible, because $\mrca(b,f)$ is in the subclade
of $d$ whereas $\mrca(a,b)$ is not.

{\em Case (iii)}.
As in case (i) we cannot have
$(c,d)$ being a parent-daughter edge and $d$ being a
``new type".
An alternate case is that 
$(c,d)$ is a parent-daughter edge and some $g \neq d$ in the subclade
of $d$ is a ``new type", and also there is some
``new type" species $f$ in the subclade determined by the
continuing edge at $c$.
By Property 1, the three species $\{f,d,g\}$ are
in different genera and their cladogram is as in Figure 4 (left side).

\setlength{\unitlength}{0.28in}
\begin{picture}(12,3.3)
\put(0,0){\line(1,1){3}}
\put(2,0){\line(1,1){1}}
\put(4,0){\line(-1,1){2}}
\put(8,0){\line(1,1){3}}
\put(10,0){\line(-1,1){1}}
\put(12,0){\line(-1,1){2}}
\put(-0.1,-0.5){f}
\put(1.8,-0.5){d}
\put(3.8,-0.5){g}
\put(7.8,-0.5){f}
\put(9.8,-0.5){c}
\put(11.8,-0.5){g}
\put(1.5,1.9){$\alpha$}
\put(9.5,1.9){$\alpha$}
\end{picture}

\vspace{0.2in}

{\bf Figure 4.}
{\small 
Cladograms arising in the proof of Theorem \ref{T1}(iii).}

\vspace{0.12in}
\noindent
As before, assume
species $b$ (which might coincide with $d$ or $g$)
is in the subclade of $d$ and species $a$ is not.
Then species $b$ must attach to the cladogram somewhere to the
lower right of $\alpha$, and species $a$ must attach to the
cladogram on one of the other edges at $\alpha$.
Whether or not the genus containing $\{a,b\}$ is 
one of the genera containing $f$ or $d$ or $g$, this
violates Property 3.

The final case is where it is the continuing edge at $c$
that is in the path from $a$ to $b$.
But in this case the same argument gives the Figure 4 (right side) cladogram;
now $a$ must attach to the branch to the lower left of $\alpha$
while $b$ must attach to one of the other two branches from $\alpha$.
Again Property 3 is violated.

\subsection{Further results for the genera classification schemes}
\label{sec-appendix}
These further results are intended to elucidate properties of the
genera classification schemes,
but (aside from Lemma \ref{Lmedchar}) are not needed for our analysis of the probability model.

Figure 1 illustrates the typical behavior of the schemes.
If one knew the true phylogenetic tree then the coarse
scheme is clearly unsatisfactory
(it puts $g$ and $r$ into the same genus despite the fact that
$g$ is more closely related to the $\{i\}$ genus than to $r$ while
$r$ is more closely related to the $\{stuv\}$ genus than to $g$).
But one can imagine settings where an unknown tree is in fact
the Figure 1 tree but, based on fragmentary fossil data, one
assigns the coarse genera.
The other two schemes seem more reasonable when one does know the 
correct tree on species.

Recall that a genus is {\em paraphyletic} if it includes
its MRCA.
Proposition \ref{P1} will show that genera produced from the
coarse scheme or the fine scheme are always 
paraphyletic,
and genera produced by the medium scheme are 
paraphyletic
except in one atypical case.
From Theorem \ref{T1} it is clear that the coarse scheme
is coarser than (or the same as) the medium scheme and the
fine scheme.  
Proposition \ref{P1}(iii) will show that, except for the same atypical case,
the medium scheme is coarser than (or the same as) the fine scheme.
Figure 5 illustrates 
what makes the case atypical: there must be some species with at least four
daughter species.

\setlength{\unitlength}{0.25in}
\begin{picture}(16,8)
\put(0,1){\line(0,1){6}}
\put(1,0){\line(0,1){2}}
\put(2,1.5){\line(0,1){1.5}}
\put(3,0){\line(0,1){4}}
\put(4,2){\line(0,1){3}}
\put(5,0){\line(0,1){6}}
\put(0,2){\line(1,0){1}}
\put(0,3){\line(1,0){2}}
\put(0,4){\line(1,0){3}}
\put(0,5){\line(1,0){4}}
\put(0,6){\line(1,0){5}}
\put(-0.2,0.6){a}
\put(0.8,-0.5){b}
\put(1.8,1.1){c}
\put(2.8,-0.5){d}
\put(3.8,1.6){e}
\put(4.8,-0.5){f}
\put(1.8,2.86){$\bullet$}
\put(2.8,3.86){$\bullet$}
\put(7,0){\line(0,1){7}}
\put(9,1.5){\line(0,1){1.5}}
\put(10,0){\line(0,1){4}}
\put(11,0){\line(0,1){5}}
\put(6.7,-0.5){ab}
\put(8.8,1.1){c}
\put(9.8,-0.5){d}
\put(10.7,-0.5){ef}
\put(7,3){\line(1,0){2}}
\put(7,4){\line(1,0){3}}
\put(7,5){\line(1,0){4}}
\put(13,0){\line(0,1){7}}
\put(15,1.5){\line(0,1){1.5}}
\put(16,0){\line(0,1){4}}
\put(12.5,-0.5){abef}
\put(14.8,1.1){c}
\put(15.8,-0.5){d}
\put(13,3){\line(1,0){2}}
\put(13,4){\line(1,0){3}}
\put(2,7){species}
\put(7.4,7){medium genera}
\put(13.1,7){fine, coarse genera}
\end{picture}

\vspace{0.2in}

{\bf Figure 5.}
{\small 
An atypical tree and its genera.
There are two ``new type" species, $\{c,d\}$.
Note that the coarse/fine genus $\{abef\}$ is
paraphyletic
while the medium genus $\{ef\}$ is not.}

\begin{Proposition}
\label{P1}
(i) Genera in the
{\bf coarse} scheme or the {\bf fine} scheme are always 
paraphyletic.
\\ (ii) 
If a {\bf medium} genus $G$ with MRCA $a$ is not 
paraphyletic,
write $(a,b)$ for the last right edge for which some species in $G$
is in the subclade of $b$.
Then subsequent to daughter $b$,
species $a$ has at least two other daughters whose subclades
contain ``new type" species.
\\ (iii)
Let $G$ be a fine genus which is not a subset of (or equal to) some medium genus.
Let $a$ be the MRCA of $G$, so $a \in G$ by part (i).
Let $b$ be the first daughter of $a$ for which the subclade of $b$ contains some
species in $G$.
Then the conclusion of part (ii) holds for this pair $(a,b)$.
\end{Proposition}
{\em Proof.}
Consider a genus $G$ whose MRCA $a$ is not in $G$, and
let $(a,b)$ be the edge specified in (ii).
The path between the leaves of some two species of $G$ must
go along the edge $(b,a)$ and upwards along the species line
of $a$ from this branchpoint $\beta$.
Because that path contains no marks, to have $a$ in a different
genus there must be a mark on the upwards path from the leaf of $a$
to the branchpoint $\beta$.
This cannot happen with the coarse or fine genera, where there are
no marks on upwards edges.
With medium genera, there cannot be a mark on the continuing
edge at $\beta$ (because $(a,b)$ has no mark).
So there must be a mark on some subsequent edge of the species $a$
line, which implies the stated conclusion in (ii).
For part (iii), let $G$ and $a$ be as in the statement.
Because $G$ is not a medium genus, there is some species $f$ in $G$ which is in a different medium genus
from $a$; let $c$ be the first daughter of $a$ such that the subclade of $c$ contains such a
species $f$.
The path from the leaf of $a$ to the leaf of $f$ contains some mark
of type (c).
 This mark must be somewhere between the leaf of $a$ and the branchpoint of edge
$(a,c)$ (otherwise there would be a mark 
of type (b) on edge $(a,c)$)
The argument in part (ii) can be repeated to obtain the stated conclusion with
$c$ in place of $b$, which implies the conclusion for $b$.

\paragraph{Numbers of marks and numbers of genera}
In the coarse and the fine schemes, each mark is on a right
edge, and the corresponding daughter species is the MRCA 
of its genus.
So by the paraphyletic property 
(Proposition \ref{P1})
in the coarse or fine scheme
the number of genera is exactly equal to the number of marks
plus one,
the ``plus one" for the genus containing the founder of the clade.
The case of medium genera is more complicated.
The path upwards from the leaf representing a species will
reach a first mark (or the founder of the whole species clade - let us add one ``virtual mark" with the founding of the clade) which does not
depend 
on choice of species in the genus,  
and which is different for different genera.
Thus each genus can be identified with a different mark. 
For instance, in Figure 5 the genus $\{ef\}$ 
is identified with the virtual mark whereas genus
$\{ab\}$
is associated with the mark on the continuing edge down
from the branchpoint of $d$ to the branchpoint of $c$.
However, not every mark has an associated genus.
For instance, if $e$ and $f$ were absent from Figure 5,
then the virtual mark would have no associated genus.
It is easy to check the following condition.
\begin{Lemma}
\label{Lmedchar}
For {\bf medium} genera, a mark is associated with a genus unless the
next downward following branchpoint is a ``rule (c)"
branchpoint, in which both parent and daughter subclade
contain new type species.

Thus the number of medium genera equals the number of marks 
(including the virtual mark)
minus the number of such marks satisfying the condition above.
\end{Lemma}

\paragraph{Operations on trees.}
Here we briefly say how the genera classifications can change when the
background structure (tree and distinguished ``new type" species)
is changed.

(a) Suppose we don't change the tree, but declare another species to be ``new type".
This can only increase the number of marks and so can only make the genera
partition become less coarse.  
For the coarse scheme it will typically create exactly one new genus.
For the other schemes it may create more than one extra genus.
For instance, in Figure 1 the designation of $s$ as a ``new type" has 
created the fine genera $\{stuv\}, \ \{opqr\}, \ \{mn\}, \ \{l\}$.
For the medium scheme it typically creates either one or two extra genera.

(b) Suppose we add an extra species to the tree (as the daughter of some already
present species) and this extra species in not ``new type".
For the fine and coarse genera, and typically for the medium genera, the new species in
put into the same genus as its parent.
Using Figure 5 we can see an atypical case with medium genera.
If a new daughter of $a$ has branchpoint between the branchpoints of $c$ and $d$,
it forms a new genus by itself, while if its branchpoint is between 
the branchpoints of $e$ and $f$ then it is put into the $\{ef\}$ genus.

\section{Tree statistics and the probability model}
A {\em statistic} of a phylogenetic tree or cladogram
is just a number (or collection of numbers)
intended to quantify some aspect of the tree.
The goal of this paper is to study,
under a probability model for 
``purely random macroevolution",
how statistics might change when one goes from
species-level trees
to trees on higher-order taxa.  That is, how statistics might change
purely as a logical consequence of the process of classification, rather than having 
any special biological significance.
For concreteness let us start by listing 
some statistics for trees,
in the setting of extant clades.
Then we state our model for species-level random macroevolution,
and finally combine ingredients to derive models of genus-level
trees.

\subsection{Examples of statistics}
\label{sec-list}
(a) Number of extant taxa.

\noindent
(b) The time series (number of taxa in existence, as a function of past time),
which includes in particular\\
\hspace*{0.3in}
 the time of origin of the tree\\
\hspace*{0.3in}
 the total number of (extinct or extant) taxa\\
\hspace*{0.3in}
 the maximum number of taxa in existence at any one time.

\noindent
(c) The time of MRCA of extant taxa\\
(d) The number of extinct taxa which are ancestral to some
extant taxon.

\noindent
(e) Statistics dealing with the shape of the cladogram on extant taxa
(see \cite{ford06,matsen06}
for recent references).

\subsection{The species-level probability model}
\label{sec-species-level}
We want a probability model for the 
phylogenetic tree on a clade with $n$ extant species 
(for given $n$) which captures the intuitive idea of
``purely random macroevolution".
Our choice is the model below, studied in detail in \cite{me111},
where some arguments (not repeated here) in its favor are presented.
In (b) the phrase ``rate $1$" means
``with probability $dt$ in each time interval of length $dt$".
So the time unit in the model equals mean species lifetime.

\paragraph{The species-level model}

\noindent
(a) The clade originates with one species at a random time before present,
whose prior distribution is uniform on $(0,\infty)$.\\
(b) As time runs forward, each species may become extinct or may speciate, each at rate $1$. \\ 
(c) Condition on the number of species at the present time $t=0$
being exactly equal to $n$.

\noindent
The ``posterior distribution" on the 
evolution of lineages given this conditioning is then
a mathematically completely defined random tree on $n$ extant species,
which we write as $c-\TREE_n$ (here $c$ is mnemonic for {\em complete})
\footnote{In (a) we use an {\em improper} [total probability
is infinite] prior distribution, but after conditioning the
posterior distribution of $c-\TREE_n$ is {\em proper} [total
probability is $1$].}.
See Figure 6 for a realization with $n = 20$.

\newpage
 \setlength{\unitlength}{0.001in}
 \begin{picture}(4000,5500)(-900,-300)
 \linethickness{0.3mm}
 \put(  3200,  54){\line(0,-1){  54}}
 \put(  1300,  1202){\line(0,-1){  1202}}
 \put(  600,  364){\line(0,-1){  364}}
 \put(  3100,  148){\line(0,-1){  148}}
 \put(  2825,  227){\line(0,-1){  227}}
 \put(  2775,  521){\line(0,-1){  521}}
 \put(  3050,  1360){\line(0,-1){  1360}}
 \put(  3175,  68){\line(0,-1){  68}}
 \put(  2800,  44){\line(0,-1){  44}}
 \put(  1575,  701){\line(0,-1){  701}}
 \put(  1250,  198){\line(0,-1){  198}}
 \put(  925,  886){\line(0,-1){  886}}
 \put(  1325,  293){\line(0,-1){  293}}
 \put(  3075,  134){\line(0,-1){  134}}
 \put(  500,  1683){\line(0,-1){  1683}}
 \put(  2850,  23){\line(0,-1){  23}}
 \put(  625,  89){\line(0,-1){  89}}
 \put(  950,  226){\line(0,-1){  226}}
 \put(  3150,  186){\line(0,-1){  186}}
 \put(  1600,  364){\line(0,-1){  364}}
 \put(  3250,  314){\line(0,-1){  292}}
 \put(  1200,  1026){\line(0,-1){  992}}
 \put(  3375,  397){\line(0,-1){  359}}
 \put(  3125,  117){\line(0,-1){  37}}
 \put(  2875,  989){\line(0,-1){  896}}
 \put(  575,  1160){\line(0,-1){  1040}}
 \put(  3225,  296){\line(0,-1){  173}}
 \put(  525,  186){\line(0,-1){  43}}
 \put(  1225,  425){\line(0,-1){  265}}
 \put(  3400,  360){\line(0,-1){  144}}
 \put(  3275,  558){\line(0,-1){  322}}
 \put(  2725,  1833){\line(0,-1){  1592}}
 \put(  3350,  1170){\line(0,-1){  920}}
 \put(  3425,  865){\line(0,-1){  605}}
 \put(  1275,  563){\line(0,-1){  303}}
 \put(  2750,  586){\line(0,-1){  249}}
 \put(  550,  368){\line(0,-1){  30}}
 \put(  3325,  814){\line(0,-1){  416}}
 \put(  1625,  1277){\line(0,-1){  709}}
 \put(  2200,  3701){\line(0,-1){  3116}}
 \put(  1550,  1611){\line(0,-1){  999}}
 \put(  2925,  1939){\line(0,-1){  1266}}
 \put(  375,  1599){\line(0,-1){  846}}
 \put(  675,  1036){\line(0,-1){  282}}
 \put(  3300,  945){\line(0,-1){  182}}
 \put(  900,  938){\line(0,-1){  65}}
 \put(  3750,  1458){\line(0,-1){  581}}
 \put(  875,  1631){\line(0,-1){  752}}
 \put(  3550,  1222){\line(0,-1){  342}}
 \put(  3825,  1259){\line(0,-1){  370}}
 \put(  3850,  1266){\line(0,-1){  374}}
 \put(  3575,  922){\line(0,-1){  18}}
 \put(  3700,  2852){\line(0,-1){  1943}}
 \put(  1500,  1475){\line(0,-1){  562}}
 \put(  650,  1180){\line(0,-1){  232}}
 \put(  1475,  1320){\line(0,-1){  367}}
 \put(  1175,  1907){\line(0,-1){  924}}
 \put(  2700,  2365){\line(0,-1){  1372}}
 \put(  3875,  1313){\line(0,-1){  298}}
 \put(  425,  1999){\line(0,-1){  906}}
 \put(  3725,  1184){\line(0,-1){  77}}
 \put(  3525,  1816){\line(0,-1){  693}}
 \put(  2100,  2048){\line(0,-1){  916}}
 \put(  450,  1194){\line(0,-1){  52}}
 \put(  1700,  2045){\line(0,-1){  902}}
 \put(  3025,  2427){\line(0,-1){  1280}}
 \put(  3500,  1698){\line(0,-1){  528}}
 \put(  3775,  1815){\line(0,-1){  639}}
 \put(  1450,  1755){\line(0,-1){  525}}
 \put(  3800,  1389){\line(0,-1){  159}}
 \put(  1400,  1991){\line(0,-1){  743}}
 \put(  2225,  2842){\line(0,-1){  1592}}
 \put(  975,  1555){\line(0,-1){  274}}
 \put(  3900,  2143){\line(0,-1){  839}}
 \put(  1375,  1767){\line(0,-1){  461}}
 \put(  2900,  1559){\line(0,-1){  249}}
 \put(  3925,  1457){\line(0,-1){  134}}
 \put(  350,  1703){\line(0,-1){  379}}
 \put(  1850,  2061){\line(0,-1){  730}}
 \put(  2950,  1413){\line(0,-1){  60}}
 \put(  1525,  1585){\line(0,-1){  220}}
 \put(  700,  1483){\line(0,-1){  105}}
 \put(  1425,  1558){\line(0,-1){  172}}
 \put(  2600,  1812){\line(0,-1){  407}}
 \put(  2500,  2604){\line(0,-1){  1183}}
 \put(  825,  3146){\line(0,-1){  1721}}
 \put(  2625,  1565){\line(0,-1){  136}}
 \put(  1350,  1578){\line(0,-1){  112}}
 \put(  475,  1551){\line(0,-1){  83}}
 \put(  2975,  1574){\line(0,-1){  90}}
 \put(  300,  1737){\line(0,-1){  225}}
 \put(  850,  1595){\line(0,-1){  44}}
 \put(  2525,  1780){\line(0,-1){  218}}
 \put(  2575,  2413){\line(0,-1){  781}}
 \put(  325,  1828){\line(0,-1){  186}}
 \put(  2075,  2712){\line(0,-1){  1037}}
 \put(  3475,  1989){\line(0,-1){  308}}
 \put(  225,  2043){\line(0,-1){  352}}
 \put(  275,  1813){\line(0,-1){  120}}
 \put(  1875,  2002){\line(0,-1){  300}}
 \put(  1650,  1747){\line(0,-1){  32}}
 \put(  2250,  2093){\line(0,-1){  377}}
 \put(  2550,  2689){\line(0,-1){  941}}
 \put(  1150,  2182){\line(0,-1){  414}}
 \put(  1675,  1997){\line(0,-1){  218}}
 \put(  250,  1887){\line(0,-1){  88}}
 \put(  400,  1889){\line(0,-1){  83}}
 \put(  1125,  3119){\line(0,-1){  1280}}
 \put(  3450,  2146){\line(0,-1){  251}}
 \put(  2650,  2060){\line(0,-1){  108}}
 \put(  725,  2042){\line(0,-1){  83}}
 \put(  750,  1994){\line(0,-1){  30}}
 \put(  150,  2815){\line(0,-1){  814}}
 \put(  200,  2081){\line(0,-1){  66}}
 \put(  175,  2185){\line(0,-1){  133}}
 \put(  2675,  2098){\line(0,-1){  44}}
 \put(  1825,  2669){\line(0,-1){  611}}
 \put(  2275,  2223){\line(0,-1){  138}}
 \put(  3000,  2697){\line(0,-1){  530}}
 \put(  2300,  2344){\line(0,-1){  154}}
 \put(  2050,  2739){\line(0,-1){  498}}
 \put(  2125,  2274){\line(0,-1){  24}}
 \put(  2475,  2704){\line(0,-1){  412}}
 \put(  1725,  2520){\line(0,-1){  89}}
 \put(  1750,  2542){\line(0,-1){  62}}
 \put(  3950,  2741){\line(0,-1){  250}}
 \put(  1900,  2532){\line(0,-1){  37}}
 \put(  2025,  2921){\line(0,-1){  422}}
 \put(  1800,  2785){\line(0,-1){  211}}
 \put(  2450,  3329){\line(0,-1){  699}}
 \put(  1075,  3948){\line(0,-1){  1289}}
 \put(  1775,  2834){\line(0,-1){  170}}
 \put(  800,  4081){\line(0,-1){  1406}}
 \put(  125,  3512){\line(0,-1){  827}}
 \put(  1950,  3433){\line(0,-1){  646}}
 \put(  1925,  2820){\line(0,-1){  9}}
 \put(  1975,  3026){\line(0,-1){  206}}
 \put(  3675,  3915){\line(0,-1){  1089}}
 \put(  2000,  2863){\line(0,-1){  24}}
 \put(  3625,  3312){\line(0,-1){  305}}
 \put(  1100,  3120){\line(0,-1){  32}}
 \put(  3600,  3410){\line(0,-1){  262}}
 \put(  2325,  3360){\line(0,-1){  155}}
 \put(  2425,  3657){\line(0,-1){  442}}
 \put(  2350,  3706){\line(0,-1){  448}}
 \put(  775,  3292){\line(0,-1){  8}}
 \put(  2375,  4105){\line(0,-1){  812}}
 \put(  2400,  3632){\line(0,-1){  288}}
 \put(  1000,  3438){\line(0,-1){  57}}
 \put(  100,  4496){\line(0,-1){  1040}}
 \put(  2175,  4567){\line(0,-1){  975}}
 \put(  3650,  3613){\line(0,-1){  2}}
 \put(  2150,  3828){\line(0,-1){  125}}
 \put(  1025,  3817){\line(0,-1){  74}}
 \put(  1050,  4039){\line(0,-1){  206}}
 \put(  75,  5490){\line(0,-1){  1157}}
 \put(  4025,  5150){\line(0,-1){  738}}
 \put(  3975,  4891){\line(0,-1){  36}}
 \put(  25,  6380){\line(0,-1){  1431}}
 \put(  4000,  5226){\line(0,-1){  121}}
 \put(  50,  5880){\line(0,-1){  416}}
 \put(  4050,  5560){\line(0,-1){  41}}
 \linethickness{0.01mm}
 \put(  3200,  54){\line(-1,0){  25}}
 \put(  1300,  1202){\line(-1,0){  125}}
 \put(  600,  364){\line(-1,0){  25}}
 \put(  3100,  148){\line(-1,0){  50}}
 \put(  2825,  227){\line(-1,0){  50}}
 \put(  2775,  521){\line(-1,0){  25}}
 \put(  3050,  1360){\line(-1,0){  25}}
 \put(  3175,  68){\line(-1,0){  25}}
 \put(  2800,  44){\line(-1,0){  25}}
 \put(  1575,  701){\line(-1,0){  25}}
 \put(  1250,  198){\line(-1,0){  25}}
 \put(  925,  886){\line(-1,0){  25}}
 \put(  1325,  293){\line(-1,0){  25}}
 \put(  3075,  134){\line(-1,0){  25}}
 \put(  500,  1683){\line(-1,0){  75}}
 \put(  2850,  23){\line(-1,0){  25}}
 \put(  625,  89){\line(-1,0){  25}}
 \put(  950,  226){\line(-1,0){  25}}
 \put(  3150,  186){\line(-1,0){  100}}
 \put(  1600,  364){\line(-1,0){  25}}
 \put(  3250,  314){\line(-1,0){  200}}
 \put(  1200,  1026){\line(-1,0){  25}}
 \put(  3375,  397){\line(-1,0){  25}}
 \put(  3125,  117){\line(-1,0){  25}}
 \put(  2875,  989){\line(-1,0){  150}}
 \put(  575,  1160){\line(-1,0){  75}}
 \put(  3225,  296){\line(-1,0){  175}}
 \put(  525,  186){\line(-1,0){  25}}
 \put(  1225,  425){\line(-1,0){  25}}
 \put(  3400,  360){\line(-1,0){  25}}
 \put(  3275,  558){\line(-1,0){  225}}
 \put(  2725,  1833){\line(-1,0){  25}}
 \put(  3350,  1170){\line(-1,0){  300}}
 \put(  3425,  865){\line(-1,0){  75}}
 \put(  1275,  563){\line(-1,0){  75}}
 \put(  2750,  586){\line(-1,0){  25}}
 \put(  550,  368){\line(-1,0){  50}}
 \put(  3325,  814){\line(-1,0){  25}}
 \put(  1625,  1277){\line(-1,0){  75}}
 \put(  2200,  3701){\line(-1,0){  25}}
 \put(  1550,  1611){\line(-1,0){  100}}
 \put(  2925,  1939){\line(-1,0){  225}}
 \put(  375,  1599){\line(-1,0){  25}}
 \put(  675,  1036){\line(-1,0){  25}}
 \put(  3300,  945){\line(-1,0){  250}}
 \put(  900,  938){\line(-1,0){  25}}
 \put(  3750,  1458){\line(-1,0){  50}}
 \put(  875,  1631){\line(-1,0){  50}}
 \put(  3550,  1222){\line(-1,0){  25}}
 \put(  3825,  1259){\line(-1,0){  25}}
 \put(  3850,  1266){\line(-1,0){  50}}
 \put(  3575,  922){\line(-1,0){  25}}
 \put(  3700,  2852){\line(-1,0){  25}}
 \put(  1500,  1475){\line(-1,0){  50}}
 \put(  650,  1180){\line(-1,0){  150}}
 \put(  1475,  1320){\line(-1,0){  25}}
 \put(  1175,  1907){\line(-1,0){  25}}
 \put(  2700,  2365){\line(-1,0){  125}}
 \put(  3875,  1313){\line(-1,0){  75}}
 \put(  425,  1999){\line(-1,0){  200}}
 \put(  3725,  1184){\line(-1,0){  25}}
 \put(  3525,  1816){\line(-1,0){  50}}
 \put(  2100,  2048){\line(-1,0){  25}}
 \put(  450,  1194){\line(-1,0){  25}}
 \put(  1700,  2045){\line(-1,0){  550}}
 \put(  3025,  2427){\line(-1,0){  25}}
 \put(  3500,  1698){\line(-1,0){  25}}
 \put(  3775,  1815){\line(-1,0){  75}}
 \put(  1450,  1755){\line(-1,0){  50}}
 \put(  3800,  1389){\line(-1,0){  25}}
 \put(  1400,  1991){\line(-1,0){  250}}
 \put(  2225,  2842){\line(-1,0){  25}}
 \put(  975,  1555){\line(-1,0){  100}}
 \put(  3900,  2143){\line(-1,0){  200}}
 \put(  1375,  1767){\line(-1,0){  200}}
 \put(  2900,  1559){\line(-1,0){  175}}
 \put(  3925,  1457){\line(-1,0){  25}}
 \put(  350,  1703){\line(-1,0){  25}}
 \put(  1850,  2061){\line(-1,0){  25}}
 \put(  2950,  1413){\line(-1,0){  25}}
 \put(  1525,  1585){\line(-1,0){  75}}
 \put(  700,  1483){\line(-1,0){  200}}
 \put(  1425,  1558){\line(-1,0){  25}}
 \put(  2600,  1812){\line(-1,0){  25}}
 \put(  2500,  2604){\line(-1,0){  25}}
 \put(  825,  3146){\line(-1,0){  25}}
 \put(  2625,  1565){\line(-1,0){  25}}
 \put(  1350,  1578){\line(-1,0){  175}}
 \put(  475,  1551){\line(-1,0){  50}}
 \put(  2975,  1574){\line(-1,0){  50}}
 \put(  300,  1737){\line(-1,0){  25}}
 \put(  850,  1595){\line(-1,0){  25}}
 \put(  2525,  1780){\line(-1,0){  25}}
 \put(  2575,  2413){\line(-1,0){  25}}
 \put(  325,  1828){\line(-1,0){  75}}
 \put(  2075,  2712){\line(-1,0){  25}}
 \put(  3475,  1989){\line(-1,0){  25}}
 \put(  225,  2043){\line(-1,0){  25}}
 \put(  275,  1813){\line(-1,0){  25}}
 \put(  1875,  2002){\line(-1,0){  25}}
 \put(  1650,  1747){\line(-1,0){  200}}
 \put(  2250,  2093){\line(-1,0){  25}}
 \put(  2550,  2689){\line(-1,0){  75}}
 \put(  1150,  2182){\line(-1,0){  25}}
 \put(  1675,  1997){\line(-1,0){  525}}
 \put(  250,  1887){\line(-1,0){  25}}
 \put(  400,  1889){\line(-1,0){  175}}
 \put(  1125,  3119){\line(-1,0){  25}}
 \put(  3450,  2146){\line(-1,0){  425}}
 \put(  2650,  2060){\line(-1,0){  75}}
 \put(  725,  2042){\line(-1,0){  500}}
 \put(  750,  1994){\line(-1,0){  25}}
 \put(  150,  2815){\line(-1,0){  25}}
 \put(  200,  2081){\line(-1,0){  25}}
 \put(  175,  2185){\line(-1,0){  25}}
 \put(  2675,  2098){\line(-1,0){  100}}
 \put(  1825,  2669){\line(-1,0){  25}}
 \put(  2275,  2223){\line(-1,0){  50}}
 \put(  3000,  2697){\line(-1,0){  525}}
 \put(  2300,  2344){\line(-1,0){  75}}
 \put(  2050,  2739){\line(-1,0){  25}}
 \put(  2125,  2274){\line(-1,0){  50}}
 \put(  2475,  2704){\line(-1,0){  25}}
 \put(  1725,  2520){\line(-1,0){  600}}
 \put(  1750,  2542){\line(-1,0){  625}}
 \put(  3950,  2741){\line(-1,0){  250}}
 \put(  1900,  2532){\line(-1,0){  75}}
 \put(  2025,  2921){\line(-1,0){  50}}
 \put(  1800,  2785){\line(-1,0){  25}}
 \put(  2450,  3329){\line(-1,0){  25}}
 \put(  1075,  3948){\line(-1,0){  25}}
 \put(  1775,  2834){\line(-1,0){  650}}
 \put(  800,  4081){\line(-1,0){  700}}
 \put(  125,  3512){\line(-1,0){  25}}
 \put(  1950,  3433){\line(-1,0){  875}}
 \put(  1925,  2820){\line(-1,0){  150}}
 \put(  1975,  3026){\line(-1,0){  25}}
 \put(  3675,  3915){\line(-1,0){  1300}}
 \put(  2000,  2863){\line(-1,0){  25}}
 \put(  3625,  3312){\line(-1,0){  25}}
 \put(  1100,  3120){\line(-1,0){  25}}
 \put(  3600,  3410){\line(-1,0){  1175}}
 \put(  2325,  3360){\line(-1,0){  125}}
 \put(  2425,  3657){\line(-1,0){  50}}
 \put(  2350,  3706){\line(-1,0){  175}}
 \put(  775,  3292){\line(-1,0){  650}}
 \put(  2375,  4105){\line(-1,0){  200}}
 \put(  2400,  3632){\line(-1,0){  25}}
 \put(  1000,  3438){\line(-1,0){  200}}
 \put(  100,  4496){\line(-1,0){  25}}
 \put(  2175,  4567){\line(-1,0){  2100}}
 \put(  3650,  3613){\line(-1,0){  1225}}
 \put(  2150,  3828){\line(-1,0){  1075}}
 \put(  1025,  3817){\line(-1,0){  225}}
 \put(  1050,  4039){\line(-1,0){  250}}
 \put(  75,  5490){\line(-1,0){  25}}
 \put(  4025,  5150){\line(-1,0){  25}}
 \put(  3975,  4891){\line(-1,0){  3900}}
 \put(  4000,  5226){\line(-1,0){  3925}}
 \put(  50,  5880){\line(-1,0){  25}}
 \put(  4050,  5560){\line(-1,0){  4000}}
 \thicklines
 \put(  -1300,  0){\line(0,1){  22}}
 \put(  -1340,  22){\line(0,1){  1}}
 \put(  -1300,  23){\line(0,1){  11}}
 \put(  -1340,  34){\line(0,1){  4}}
 \put(  -1380,  38){\line(0,1){  6}}
 \put(  -1340,  44){\line(0,1){  10}}
 \put(  -1300,  54){\line(0,1){  14}}
 \put(  -1260,  68){\line(0,1){  12}}
 \put(  -1300,  80){\line(0,1){  9}}
 \put(  -1260,  89){\line(0,1){  4}}
 \put(  -1300,  93){\line(0,1){  24}}
 \put(  -1260,  117){\line(0,1){  3}}
 \put(  -1300,  120){\line(0,1){  3}}
 \put(  -1340,  123){\line(0,1){  11}}
 \put(  -1300,  134){\line(0,1){  9}}
 \put(  -1340,  143){\line(0,1){  5}}
 \put(  -1300,  148){\line(0,1){  12}}
 \put(  -1340,  160){\line(0,1){  26}}
 \put(  -1300,  186){\line(0,1){  0}}
 \put(  -1260,  186){\line(0,1){  12}}
 \put(  -1220,  198){\line(0,1){  18}}
 \put(  -1260,  216){\line(0,1){  10}}
 \put(  -1220,  226){\line(0,1){  1}}
 \put(  -1180,  227){\line(0,1){  9}}
 \put(  -1220,  236){\line(0,1){  5}}
 \put(  -1260,  241){\line(0,1){  9}}
 \put(  -1300,  250){\line(0,1){  10}}
 \put(  -1340,  260){\line(0,1){  0}}
 \put(  -1380,  260){\line(0,1){  33}}
 \put(  -1340,  293){\line(0,1){  3}}
 \put(  -1300,  296){\line(0,1){  18}}
 \put(  -1260,  314){\line(0,1){  23}}
 \put(  -1300,  337){\line(0,1){  1}}
 \put(  -1340,  338){\line(0,1){  22}}
 \put(  -1300,  360){\line(0,1){  4}}
 \put(  -1260,  364){\line(0,1){  0}}
 \put(  -1220,  364){\line(0,1){  4}}
 \put(  -1180,  368){\line(0,1){  29}}
 \put(  -1140,  397){\line(0,1){  1}}
 \put(  -1180,  398){\line(0,1){  27}}
 \put(  -1140,  425){\line(0,1){  96}}
 \put(  -1100,  521){\line(0,1){  37}}
 \put(  -1060,  558){\line(0,1){  5}}
 \put(  -1020,  563){\line(0,1){  5}}
 \put(  -1060,  568){\line(0,1){  17}}
 \put(  -1100,  585){\line(0,1){  1}}
 \put(  -1060,  586){\line(0,1){  26}}
 \put(  -1100,  612){\line(0,1){  61}}
 \put(  -1140,  673){\line(0,1){  28}}
 \put(  -1100,  701){\line(0,1){  52}}
 \put(  -1140,  753){\line(0,1){  1}}
 \put(  -1180,  754){\line(0,1){  9}}
 \put(  -1220,  763){\line(0,1){  51}}
 \put(  -1180,  814){\line(0,1){  51}}
 \put(  -1140,  865){\line(0,1){  8}}
 \put(  -1180,  873){\line(0,1){  4}}
 \put(  -1220,  877){\line(0,1){  2}}
 \put(  -1260,  879){\line(0,1){  1}}
 \put(  -1300,  880){\line(0,1){  6}}
 \put(  -1260,  886){\line(0,1){  3}}
 \put(  -1300,  889){\line(0,1){  3}}
 \put(  -1340,  892){\line(0,1){  12}}
 \put(  -1380,  904){\line(0,1){  5}}
 \put(  -1420,  909){\line(0,1){  4}}
 \put(  -1460,  913){\line(0,1){  9}}
 \put(  -1420,  922){\line(0,1){  16}}
 \put(  -1380,  938){\line(0,1){  7}}
 \put(  -1340,  945){\line(0,1){  3}}
 \put(  -1380,  948){\line(0,1){  5}}
 \put(  -1420,  953){\line(0,1){  30}}
 \put(  -1460,  983){\line(0,1){  6}}
 \put(  -1420,  989){\line(0,1){  4}}
 \put(  -1460,  993){\line(0,1){  22}}
 \put(  -1500,  1015){\line(0,1){  11}}
 \put(  -1460,  1026){\line(0,1){  10}}
 \put(  -1420,  1036){\line(0,1){  57}}
 \put(  -1460,  1093){\line(0,1){  14}}
 \put(  -1500,  1107){\line(0,1){  16}}
 \put(  -1540,  1123){\line(0,1){  9}}
 \put(  -1580,  1132){\line(0,1){  10}}
 \put(  -1620,  1142){\line(0,1){  1}}
 \put(  -1660,  1143){\line(0,1){  4}}
 \put(  -1700,  1147){\line(0,1){  13}}
 \put(  -1660,  1160){\line(0,1){  10}}
 \put(  -1700,  1170){\line(0,1){  0}}
 \put(  -1660,  1170){\line(0,1){  6}}
 \put(  -1700,  1176){\line(0,1){  4}}
 \put(  -1660,  1180){\line(0,1){  4}}
 \put(  -1620,  1184){\line(0,1){  10}}
 \put(  -1580,  1194){\line(0,1){  8}}
 \put(  -1540,  1202){\line(0,1){  20}}
 \put(  -1500,  1222){\line(0,1){  8}}
 \put(  -1540,  1230){\line(0,1){  0}}
 \put(  -1580,  1230){\line(0,1){  18}}
 \put(  -1620,  1248){\line(0,1){  2}}
 \put(  -1660,  1250){\line(0,1){  9}}
 \put(  -1620,  1259){\line(0,1){  7}}
 \put(  -1580,  1266){\line(0,1){  11}}
 \put(  -1540,  1277){\line(0,1){  4}}
 \put(  -1580,  1281){\line(0,1){  23}}
 \put(  -1620,  1304){\line(0,1){  2}}
 \put(  -1660,  1306){\line(0,1){  4}}
 \put(  -1700,  1310){\line(0,1){  3}}
 \put(  -1660,  1313){\line(0,1){  7}}
 \put(  -1620,  1320){\line(0,1){  3}}
 \put(  -1660,  1323){\line(0,1){  1}}
 \put(  -1700,  1324){\line(0,1){  7}}
 \put(  -1740,  1331){\line(0,1){  22}}
 \put(  -1780,  1353){\line(0,1){  7}}
 \put(  -1740,  1360){\line(0,1){  5}}
 \put(  -1780,  1365){\line(0,1){  13}}
 \put(  -1820,  1378){\line(0,1){  8}}
 \put(  -1860,  1386){\line(0,1){  3}}
 \put(  -1820,  1389){\line(0,1){  16}}
 \put(  -1860,  1405){\line(0,1){  8}}
 \put(  -1820,  1413){\line(0,1){  8}}
 \put(  -1860,  1421){\line(0,1){  4}}
 \put(  -1900,  1425){\line(0,1){  4}}
 \put(  -1940,  1429){\line(0,1){  28}}
 \put(  -1900,  1457){\line(0,1){  1}}
 \put(  -1860,  1458){\line(0,1){  8}}
 \put(  -1900,  1466){\line(0,1){  2}}
 \put(  -1940,  1468){\line(0,1){  7}}
 \put(  -1900,  1475){\line(0,1){  8}}
 \put(  -1860,  1483){\line(0,1){  1}}
 \put(  -1900,  1484){\line(0,1){  28}}
 \put(  -1940,  1512){\line(0,1){  39}}
 \put(  -1900,  1551){\line(0,1){  0}}
 \put(  -1940,  1551){\line(0,1){  4}}
 \put(  -1900,  1555){\line(0,1){  3}}
 \put(  -1860,  1558){\line(0,1){  1}}
 \put(  -1820,  1559){\line(0,1){  3}}
 \put(  -1860,  1562){\line(0,1){  3}}
 \put(  -1820,  1565){\line(0,1){  9}}
 \put(  -1780,  1574){\line(0,1){  4}}
 \put(  -1740,  1578){\line(0,1){  7}}
 \put(  -1700,  1585){\line(0,1){  10}}
 \put(  -1660,  1595){\line(0,1){  4}}
 \put(  -1620,  1599){\line(0,1){  12}}
 \put(  -1580,  1611){\line(0,1){  20}}
 \put(  -1540,  1631){\line(0,1){  1}}
 \put(  -1580,  1632){\line(0,1){  10}}
 \put(  -1620,  1642){\line(0,1){  33}}
 \put(  -1660,  1675){\line(0,1){  6}}
 \put(  -1700,  1681){\line(0,1){  2}}
 \put(  -1660,  1683){\line(0,1){  8}}
 \put(  -1700,  1691){\line(0,1){  2}}
 \put(  -1740,  1693){\line(0,1){  5}}
 \put(  -1700,  1698){\line(0,1){  4}}
 \put(  -1740,  1702){\line(0,1){  1}}
 \put(  -1700,  1703){\line(0,1){  12}}
 \put(  -1740,  1715){\line(0,1){  1}}
 \put(  -1780,  1716){\line(0,1){  21}}
 \put(  -1740,  1737){\line(0,1){  10}}
 \put(  -1700,  1747){\line(0,1){  1}}
 \put(  -1740,  1748){\line(0,1){  7}}
 \put(  -1700,  1755){\line(0,1){  12}}
 \put(  -1660,  1767){\line(0,1){  1}}
 \put(  -1700,  1768){\line(0,1){  11}}
 \put(  -1740,  1779){\line(0,1){  1}}
 \put(  -1700,  1780){\line(0,1){  19}}
 \put(  -1740,  1799){\line(0,1){  7}}
 \put(  -1780,  1806){\line(0,1){  6}}
 \put(  -1740,  1812){\line(0,1){  1}}
 \put(  -1700,  1813){\line(0,1){  2}}
 \put(  -1660,  1815){\line(0,1){  1}}
 \put(  -1620,  1816){\line(0,1){  12}}
 \put(  -1580,  1828){\line(0,1){  5}}
 \put(  -1540,  1833){\line(0,1){  6}}
 \put(  -1580,  1839){\line(0,1){  48}}
 \put(  -1540,  1887){\line(0,1){  2}}
 \put(  -1500,  1889){\line(0,1){  6}}
 \put(  -1540,  1895){\line(0,1){  12}}
 \put(  -1500,  1907){\line(0,1){  32}}
 \put(  -1460,  1939){\line(0,1){  13}}
 \put(  -1500,  1952){\line(0,1){  7}}
 \put(  -1540,  1959){\line(0,1){  5}}
 \put(  -1580,  1964){\line(0,1){  25}}
 \put(  -1540,  1989){\line(0,1){  2}}
 \put(  -1500,  1991){\line(0,1){  3}}
 \put(  -1460,  1994){\line(0,1){  3}}
 \put(  -1420,  1997){\line(0,1){  2}}
 \put(  -1380,  1999){\line(0,1){  2}}
 \put(  -1420,  2001){\line(0,1){  1}}
 \put(  -1380,  2002){\line(0,1){  13}}
 \put(  -1420,  2015){\line(0,1){  27}}
 \put(  -1380,  2042){\line(0,1){  1}}
 \put(  -1340,  2043){\line(0,1){  2}}
 \put(  -1300,  2045){\line(0,1){  3}}
 \put(  -1260,  2048){\line(0,1){  4}}
 \put(  -1300,  2052){\line(0,1){  2}}
 \put(  -1340,  2054){\line(0,1){  4}}
 \put(  -1380,  2058){\line(0,1){  2}}
 \put(  -1340,  2060){\line(0,1){  1}}
 \put(  -1300,  2061){\line(0,1){  20}}
 \put(  -1260,  2081){\line(0,1){  4}}
 \put(  -1300,  2085){\line(0,1){  8}}
 \put(  -1260,  2093){\line(0,1){  5}}
 \put(  -1220,  2098){\line(0,1){  45}}
 \put(  -1180,  2143){\line(0,1){  3}}
 \put(  -1140,  2146){\line(0,1){  21}}
 \put(  -1180,  2167){\line(0,1){  15}}
 \put(  -1140,  2182){\line(0,1){  3}}
 \put(  -1100,  2185){\line(0,1){  5}}
 \put(  -1140,  2190){\line(0,1){  33}}
 \put(  -1100,  2223){\line(0,1){  18}}
 \put(  -1140,  2241){\line(0,1){  9}}
 \put(  -1180,  2250){\line(0,1){  24}}
 \put(  -1140,  2274){\line(0,1){  18}}
 \put(  -1180,  2292){\line(0,1){  52}}
 \put(  -1140,  2344){\line(0,1){  21}}
 \put(  -1100,  2365){\line(0,1){  48}}
 \put(  -1060,  2413){\line(0,1){  14}}
 \put(  -1020,  2427){\line(0,1){  4}}
 \put(  -1060,  2431){\line(0,1){  49}}
 \put(  -1100,  2480){\line(0,1){  11}}
 \put(  -1140,  2491){\line(0,1){  4}}
 \put(  -1180,  2495){\line(0,1){  4}}
 \put(  -1220,  2499){\line(0,1){  21}}
 \put(  -1180,  2520){\line(0,1){  12}}
 \put(  -1140,  2532){\line(0,1){  10}}
 \put(  -1100,  2542){\line(0,1){  32}}
 \put(  -1140,  2574){\line(0,1){  30}}
 \put(  -1100,  2604){\line(0,1){  26}}
 \put(  -1140,  2630){\line(0,1){  29}}
 \put(  -1180,  2659){\line(0,1){  5}}
 \put(  -1220,  2664){\line(0,1){  5}}
 \put(  -1180,  2669){\line(0,1){  6}}
 \put(  -1220,  2675){\line(0,1){  10}}
 \put(  -1260,  2685){\line(0,1){  4}}
 \put(  -1220,  2689){\line(0,1){  8}}
 \put(  -1180,  2697){\line(0,1){  7}}
 \put(  -1140,  2704){\line(0,1){  8}}
 \put(  -1100,  2712){\line(0,1){  27}}
 \put(  -1060,  2739){\line(0,1){  2}}
 \put(  -1020,  2741){\line(0,1){  44}}
 \put(  -980,  2785){\line(0,1){  2}}
 \put(  -1020,  2787){\line(0,1){  24}}
 \put(  -1060,  2811){\line(0,1){  4}}
 \put(  -1020,  2815){\line(0,1){  5}}
 \put(  -980,  2820){\line(0,1){  0}}
 \put(  -1020,  2820){\line(0,1){  6}}
 \put(  -1060,  2826){\line(0,1){  8}}
 \put(  -1020,  2834){\line(0,1){  5}}
 \put(  -1060,  2839){\line(0,1){  3}}
 \put(  -1020,  2842){\line(0,1){  10}}
 \put(  -980,  2852){\line(0,1){  11}}
 \put(  -940,  2863){\line(0,1){  58}}
 \put(  -900,  2921){\line(0,1){  86}}
 \put(  -940,  3007){\line(0,1){  19}}
 \put(  -900,  3026){\line(0,1){  62}}
 \put(  -940,  3088){\line(0,1){  31}}
 \put(  -900,  3119){\line(0,1){  1}}
 \put(  -860,  3120){\line(0,1){  26}}
 \put(  -820,  3146){\line(0,1){  2}}
 \put(  -860,  3148){\line(0,1){  57}}
 \put(  -900,  3205){\line(0,1){  10}}
 \put(  -940,  3215){\line(0,1){  43}}
 \put(  -980,  3258){\line(0,1){  26}}
 \put(  -1020,  3284){\line(0,1){  8}}
 \put(  -980,  3292){\line(0,1){  1}}
 \put(  -1020,  3293){\line(0,1){  19}}
 \put(  -980,  3312){\line(0,1){  17}}
 \put(  -940,  3329){\line(0,1){  15}}
 \put(  -980,  3344){\line(0,1){  16}}
 \put(  -940,  3360){\line(0,1){  21}}
 \put(  -980,  3381){\line(0,1){  29}}
 \put(  -940,  3410){\line(0,1){  23}}
 \put(  -900,  3433){\line(0,1){  5}}
 \put(  -860,  3438){\line(0,1){  18}}
 \put(  -900,  3456){\line(0,1){  56}}
 \put(  -860,  3512){\line(0,1){  80}}
 \put(  -900,  3592){\line(0,1){  19}}
 \put(  -940,  3611){\line(0,1){  2}}
 \put(  -900,  3613){\line(0,1){  19}}
 \put(  -860,  3632){\line(0,1){  25}}
 \put(  -820,  3657){\line(0,1){  44}}
 \put(  -780,  3701){\line(0,1){  2}}
 \put(  -820,  3703){\line(0,1){  3}}
 \put(  -780,  3706){\line(0,1){  37}}
 \put(  -820,  3743){\line(0,1){  74}}
 \put(  -780,  3817){\line(0,1){  11}}
 \put(  -740,  3828){\line(0,1){  5}}
 \put(  -780,  3833){\line(0,1){  82}}
 \put(  -740,  3915){\line(0,1){  33}}
 \put(  -700,  3948){\line(0,1){  91}}
 \put(  -660,  4039){\line(0,1){  42}}
 \put(  -620,  4081){\line(0,1){  24}}
 \put(  -580,  4105){\line(0,1){  228}}
 \put(  -620,  4333){\line(0,1){  79}}
 \put(  -660,  4412){\line(0,1){  84}}
 \put(  -620,  4496){\line(0,1){  71}}
 \put(  -580,  4567){\line(0,1){  288}}
 \put(  -620,  4855){\line(0,1){  36}}
 \put(  -580,  4891){\line(0,1){  58}}
 \put(  -620,  4949){\line(0,1){  156}}
 \put(  -660,  5105){\line(0,1){  45}}
 \put(  -620,  5150){\line(0,1){  76}}
 \put(  -580,  5226){\line(0,1){  238}}
 \put(  -620,  5464){\line(0,1){  26}}
 \put(  -580,  5490){\line(0,1){  29}}
 \put(  -620,  5519){\line(0,1){  41}}
 \put(  -580,  5560){\line(0,1){  320}}
 \put(  -540,  5880){\line(0,1){  500}}
 \thinlines
 \put(-500,0){\line(0,1){  6380}}
 \put(  -1300,  0){\line(0,1){  23}}
 \put(  -1260,  23){\line(0,1){  21}}
 \put(  -1220,  44){\line(0,1){  10}}
 \put(  -1180,  54){\line(0,1){  14}}
 \put(  -1140,  68){\line(0,1){  21}}
 \put(  -1100,  89){\line(0,1){  45}}
 \put(  -1060,  134){\line(0,1){  14}}
 \put(  -1020,  148){\line(0,1){  38}}
 \put(  -980,  186){\line(0,1){  40}}
 \put(  -940,  226){\line(0,1){  1}}
 \put(  -900,  227){\line(0,1){  66}}
 \put(  -860,  293){\line(0,1){  71}}
 \put(  -820,  364){\line(0,1){  796}}
 \put(  -780,  1160){\line(0,1){  42}}
 \put(  -740,  1202){\line(0,1){  789}}
 \put(  -700,  1991){\line(0,1){  706}}
 \put(  -660,  2697){\line(0,1){  1342}}
 \put(  -620,  4039){\line(0,1){  42}}
 \put(  -580,  4081){\line(0,1){  486}}
 \put(  -540,  4567){\line(0,1){  1813}}
 \put(-500,0){\line(1,0){50}}
 \put(-500,1000){\line(1,0){50}}
 \put(-500,2000){\line(1,0){50}}
 \put(-500,3000){\line(1,0){50}}
 \put(-500,4000){\line(1,0){50}}
 \put(-500,5000){\line(1,0){50}}
 \put(-540,0){\line(-1,0){1560}}
 \put(-540,0){\line(0,-1){50}}
 \put(-1300,0){\line(0,-1){50}}
 \put(-2100,0){\line(0,-1){50}}
 \put(-410,-43){0}
 \put(-410,957){2}
 \put(-420,680){time}
 \put(-420,510){(b.p.)}
 \put(-410,1957){4}
 \put(-410,2957){6}
 \put(-410,3957){8}
 \put(-410,4957){10}
 \put(-500,  6380){\line(1,0){50}}
 \put(-410,  6337){$\Torigin$}
 \put(-500,  4567){\line(1,0){50}}
 \put(-410,  4524){$\Tmrca$}
 \put(-570,-150){1}
 \put(-1350,-150){20}
 \put(-2150,-150){40}
 \put(-1900,-330){number of species}
\put(3351,1172){$\bullet$}
\put(976,1557){$\bullet$}
\put(201,2075){$\bullet$}
\put(251,1889){$\bullet$}
\put(751,1996){$\bullet$}
\put(2376,4107){$\bullet$}
\put(76,5492){$\bullet$}
 \end{picture}

\vspace{0.2in}

{\bf Figure 6.}
{\small
A realization of $c-\TREE_{20}$, a complete clade on $20$
extant species.
The figure is drawn so that each species occupies a vertical line
(from time of origin to time of extinction (or present)), different
species evenly
spaced left-to-right (so that each subclade is a consecutive series),
using the convention:
daughters are to right of parents, earlier daughters rightmost.
On the left are time series: 
the outer line is total number of species, the inner line is
number of ancestors of extant species.
Marks $\bullet$ indicate ``new type" species, used later to construct genera.
In this realization there were a total number $142$
of extinct species,
with a maximum of $38$ species at one time;
$\Tmrca = 9.05$
and $\Torigin = 12.75$.
}

\newpage

This model is sufficiently simple that one can do many calculations
(exact formulas for given $n$, and $n \to \infty$ asymptotic
approximations). 
See \cite{me111} for results for the phylogenetic tree statistics
mentioned in section \ref{sec-list}.
The induced cladogram on the $n$ extant species 
has the same distribution
(ERM, for {\em equal rates Markov})
as in simpler models 
such as the Yule process (speciations but no extinctions: see next section) or the
Moran/coalescent models (number of species fixed at $n$,
with simultaneous extinction/speciation of two random species).
Properties of this distribution are well understood
\cite{me93}.

One lesson from \cite{me111} is that in this model, many
phylogenetic tree statistics are highly variable between
realizations.  
For instance, the time since MRCA scales as 
$nT$ where $T$ is a random variable with
mean $1$ but with infinite variance.
This phenomenon is rather hidden in our formulas but will be
re-emphasized in the sequel 
\cite{me-macro}.

\subsection{The probability model for higher-order taxa}
\label{sec-pmhot}
Start with the model above for
the complete tree $c-\TREE_n$ on a clade with $n$ extant
species.  Introduce a parameter
$0<\theta <1$,
and suppose that each species (extinct or extant)
independently
has chance $\theta$ to be a new type.
Then any of the three schemes from section \ref{sec-3g} can be
used to 
define an induced tree on genera, which we shall call
$\GENERA^{\theta, \mbox{\tiny fine}}_{n\,\mbox{\tiny  species}}$
or
$\GENERA^{\theta, \mbox{\tiny medium}}_{n\,\mbox{\tiny  species}}$
or
$\GENERA^{\theta, \mbox{\tiny coarse}}_{n\,\mbox{\tiny  species}}$.
Figures 7 and 8 show realizations derived from the 162-species clade
in Figure 6.
See our web site
{\tt www.stat.berkeley.edu/users/aldous/Research/Phylo/index.html} 
for further realizations.
Decreasing the parameter $\theta$ will increase the average number
of species per genus:
alternatively, regard decreasing $\theta$ as moving up the
taxonomic hierarchy.

This framework for probability models of higher-order macroevolution
is the conceptual novelty of this paper, so 
(before proceeding to mathematical calculations in the next section) 
let us add some discussion.

\newpage
 \setlength{\unitlength}{0.001in}
 \begin{picture}(4000,5500)(-300,0)
 \put(  137,  6380){\line(0,-1){  1430}}
\put(122,-127){a}
 \put(  274,  5880){\line(0,-1){  416}}
\put(259,-127){a}
 \put(  411,  5490){\line(0,-1){  1156}}
\put(396,-127){b}
 \put(  548,  4496){\line(0,-1){  1039}}
\put(533,-127){c}
 \put(  685,  3512){\line(0,-1){  826}}
\put(670,-127){c}
 \put(  822,  2815){\line(0,-1){  813}}
\put(807,-127){c}
 \put(  959,  2185){\line(0,-1){  133}}
\put(944,-127){c}
 \put(  1096,  2081){\line(0,-1){  66}}
\put(1081,-127){d}
 \put(  1233,  2043){\line(0,-1){  2043}}
\put(1218,-127){e}
 \put(  1370,  1887){\line(0,-1){  1133}}
\put(1355,-127){f}
 \put(  1507,  2042){\line(0,-1){  83}}
\put(1492,-127){g}
 \put(  1644,  1994){\line(0,-1){  29}}
\put(1629,-127){h}
 \put(  1781,  4081){\line(0,-1){  4081}}
\put(1766,-127){i}
 \put(  1918,  3146){\line(0,-1){  1721}}
\put(1903,-127){i}
 \put(  2055,  1631){\line(0,-1){  1631}}
\put(2040,-127){i}
 \put(  2192,  1555){\line(0,-1){  274}}
\put(2177,-127){j}
 \put(  2329,  4567){\line(0,-1){  3981}}
\put(2314,-127){k}
 \put(  2466,  4105){\line(0,-1){  3228}}
\put(2451,-127){l}
 \put(  2603,  3657){\line(0,-1){  650}}
\put(2588,-127){l}
 \put(  2740,  3329){\line(0,-1){  698}}
\put(2725,-127){l}
 \put(  2877,  2704){\line(0,-1){  2704}}
\put(2862,-127){l}
 \put(  3014,  2697){\line(0,-1){  530}}
\put(2999,-127){l}
 \put(  3151,  2427){\line(0,-1){  1547}}
\put(3136,-127){l}
 \put(  3288,  1360){\line(0,-1){  1360}}
\put(3273,-127){l}
 \put(  3425,  1170){\line(0,-1){  1131}}
\put(3410,-127){m}
 \put(  274,  5880){\line(-1,0){  137}}
 \put(  411,  5490){\line(-1,0){  137}}
 \put(  548,  4496){\line(-1,0){  137}}
 \put(  685,  3512){\line(-1,0){  137}}
 \put(  822,  2815){\line(-1,0){  137}}
 \put(  959,  2185){\line(-1,0){  137}}
 \put(  1096,  2081){\line(-1,0){  137}}
 \put(  1233,  2043){\line(-1,0){  137}}
 \put(  1370,  1887){\line(-1,0){  137}}
 \put(  1507,  2042){\line(-1,0){  274}}
 \put(  1644,  1994){\line(-1,0){  137}}
 \put(  1781,  4081){\line(-1,0){  1233}}
 \put(  1918,  3146){\line(-1,0){  137}}
 \put(  2055,  1631){\line(-1,0){  137}}
 \put(  2192,  1555){\line(-1,0){  137}}
 \put(  2329,  4567){\line(-1,0){  1918}}
 \put(  2466,  4105){\line(-1,0){  137}}
 \put(  2603,  3657){\line(-1,0){  137}}
 \put(  2740,  3329){\line(-1,0){  137}}
 \put(  2877,  2704){\line(-1,0){  137}}
 \put(  3014,  2697){\line(-1,0){  137}}
 \put(  3151,  2427){\line(-1,0){  137}}
 \put(  3288,  1360){\line(-1,0){  137}}
 \put(  3425,  1170){\line(-1,0){  137}}
 \thinlines
 \put(-500,0){\line(0,1){  6380}}
 \put(-500,0){\line(1,0){50}}
 \put(-500,1000){\line(1,0){50}}
 \put(-500,2000){\line(1,0){50}}
 \put(-500,3000){\line(1,0){50}}
 \put(-500,4000){\line(1,0){50}}
 \put(-500,5000){\line(1,0){50}}
 \put(-410,-43){0}
 \put(-410,957){2}
 \put(-420,680){time}
 \put(-420,510){(b.p.)}
 \put(-410,1957){4}
 \put(-410,2957){6}
 \put(-410,3957){8}
 \put(-410,4957){10}
 \put(-500,  6380){\line(1,0){50}}
 \put(-410,  6337){$\Torigin_{20}$}
 \put(-500,  4567){\line(1,0){50}}
 \put(-410,  4524){$\Tmrca_{20}$}
\put(411,5492){$\bullet$}
\put(1096,2079){$\bullet$}
\put(1370,1889){$\bullet$}
\put(1644,1996){$\bullet$}
\put(2192,1557){$\bullet$}
\put(2466,4107){$\bullet$}
\put(3425,1172){$\bullet$}
 \end{picture}

\vspace{0.4in}
{\bf Figure 7.}
{\small 
A realization of the tree 
$\GENERA^{\theta, \mbox{\tiny fine}}_{n\,\mbox{\tiny  species}}$
on extant and extinct fine genera,
with $n = 20$ extant species and $\theta = 0.04$.
It was derived from the realization of $c-\TREE_{20}$ in Figure 1,
with the ``new type" species there indicated by $\bullet$.
In this realization, there were $7$ such ``new type" species,
producing $25$ genera, of which $5$ were extant.
Letters $\{$a, b, c, \ldots ,m $\}$ indicate which of these fine genera are combined to
form medium genera in Figure 8.
}

\newpage
 \setlength{\unitlength}{0.0006in}
 \begin{picture}(4000,5500)(-300,-100)
 \put(  137,  6380){\line(0,-1){  1430}}
\put(122,-197){a}
 \put(  411,  5490){\line(0,-1){  1156}}
\put(396,-197){b}
 \put(  685,  4496){\line(0,-1){  2444}}
\put(670,-197){c}
 \put(  959,  2081){\line(0,-1){  66}}
\put(944,-197){d}
 \put(  1233,  2043){\line(0,-1){  2043}}
\put(1218,-197){e}
 \put(  1507,  1887){\line(0,-1){  1133}}
\put(1492,-197){f}
 \put(  1781,  2042){\line(0,-1){  83}}
\put(1766,-197){g}
 \put(  2055,  1994){\line(0,-1){  29}}
\put(2040,-197){h}
 \put(  2329,  4081){\line(0,-1){  4081}}
\put(2314,-197){i}
 \put(  2603,  1555){\line(0,-1){  274}}
\put(2588,-197){j}
 \put(  2877,  4567){\line(0,-1){  3981}}
\put(2862,-197){k}
 \put(  3151,  4105){\line(0,-1){  4105}}
\put(3136,-197){l}
 \put(  3425,  1170){\line(0,-1){  1131}}
\put(3410,-197){m}
 \put(  411,  5490){\line(-1,0){  274}}
 \put(  685,  4496){\line(-1,0){  274}}
 \put(  959,  2081){\line(-1,0){  274}}
 \put(  1233,  2043){\line(-1,0){  274}}
 \put(  1507,  1887){\line(-1,0){  274}}
 \put(  1781,  2042){\line(-1,0){  548}}
 \put(  2055,  1994){\line(-1,0){  274}}
 \put(  2329,  4081){\line(-1,0){  1644}}
 \put(  2603,  1555){\line(-1,0){  274}}
 \put(  2877,  4567){\line(-1,0){  2466}}
 \put(  3151,  4105){\line(-1,0){  274}}
 \put(  3425,  1170){\line(-1,0){  274}}
\put(411,5492){$\bullet$}
\put(949,2079){$\bullet$}
\put(1517,1881){$\bullet$}
\put(2055,1996){$\bullet$}
\put(2618,1557){$\bullet$}
\put(3166,4107){$\bullet$}
\put(3425,1172){$\bullet$}
 \thinlines
 \put(-800,0){\line(0,1){  6380}}
 \put(-800,0){\line(1,0){50}}
 \put(-800,1000){\line(1,0){50}}
 \put(-800,2000){\line(1,0){50}}
 \put(-800,3000){\line(1,0){50}}
 \put(-800,4000){\line(1,0){50}}
 \put(-800,5000){\line(1,0){50}}
 \put(-710,-43){0}
 \put(-710,957){2}
 \put(-720,680){time}
 \put(-720,510){(b.p.)}
 \put(-710,1957){4}
 \put(-710,2957){6}
 \put(-710,3957){8}
 \put(-710,4957){10}
 \put(-800,  6380){\line(1,0){50}}
 \put(-710,  6337){$\Torigin_{20}$}
 \put(-800,  4567){\line(1,0){50}}
 \put(-710,  4524){$\Tmrca_{20}$}
 \put(  4137,  6380){\line(0,-1){  1430}}
\put(4122,-197){a}
 \put(  4411,  5490){\line(0,-1){  5490}}
\put(4396,-197){b}
\put(4396,-377){c}
\put(4396,-577){i}
\put(4396,-777){k}
 \put(  4685,  1555){\line(0,-1){  274}}
\put(4685,-197){j}
 \put(  4959,  2081){\line(0,-1){  2081}}
\put(4959,-197){d}
\put(4959,-377){e}
\put(4959,-577){g}
 \put(  5233,  1887){\line(0,-1){  1133}}
\put(5218,-197){f}
 \put(  5507,  1994){\line(0,-1){  29}}
\put(5492,-197){h}
 \put(  5781,  4105){\line(0,-1){  4105}}
\put(5766,-197){l}
 \put(  6055,  1170){\line(0,-1){  1131}}
\put(6040,-197){m}
 \put(  4411,  5490){\line(-1,0){  274}}
 \put(  4685,  1555){\line(-1,0){  274}}
 \put(  4959,  2081){\line(-1,0){  548}}
 \put(  5233,  1887){\line(-1,0){  274}}
 \put(  5507,  1994){\line(-1,0){  548}}
 \put(  5781,  4105){\line(-1,0){  1370}}
 \put(  6055,  1170){\line(-1,0){  274}}
\put(4411,5492){$\bullet$}
\put(4974,2079){$\bullet$}
\put(5233,1881){$\bullet$}
\put(5507,1996){$\bullet$}
\put(4700,1557){$\bullet$}
\put(5781,4107){$\bullet$}
\put(6055,1172){$\bullet$}
\end{picture}

\vspace{0.4in}
{\bf Figure 8.}
{\small 
The phylogenetic trees on medium genera (left) and coarse genera (right)
corresponding to the fine genera in Figure 7.
There are 13 medium and 8 coarse genera.
}

\paragraph{Ingredients of model}
We can view the model as having three ingredients:
\begin{itemize}
\item 
the probability model (section \ref{sec-species-level})
for phylogenetic trees on species
\item 
the idea of using ``new type" species to define genera,
and the probability model above for new type species
\item 
the particular classification schemes for defining genera.
\end{itemize}
Obviously one could choose to vary details of the first and third ingredients.

\paragraph{Previous models}
Yule \cite{yule24} proposed the basic model for speciations
without extinctions.
Initially there is one species; 
each species has daughter species at rate $1$.
Though this species model is familiar nowadays, the main point
of Yule's work is invariably overlooked.
He superimposed a model of {\em genera} by supposing that,
from within each existing genus, a new species of new genus
arises at some constant stochastic rate $\lambda$.
This leads to a one-parameter family of long-tailed distributions
for number of species per genus 
(see \cite{me93} for brief description).
Yule's model perhaps foreshadows 
``hierarchical selection above the species level"
\cite{gouldSET};
in contrast, our model for higher-level taxa 
does not assume separate genus-level biological effect, but rather
combines species-level novelty with conventions about how
systematicians construct genera.
In other words,
our model is intended to capture the ``neutral" idea that
a subclade is defined by a heritable character but that this
character has no ``selective advantage", 
i.e. that the species in the subclade have unchanged speciation and
extinction rates.
This ``neutral" idea was studied in
\cite{gouldSE}, 
but there the partition of species into genera was based only 
on the size of subclades.

\paragraph{Nuances of the model}
We claim that the model is sufficiently flexible that any question
one might ask about genera-level macroevolution statistics can be formulated
within the model.
Our initial description assumed a clade with a specified number $n$
of species (and hence a random number of genera), but it's usually more natural to work in one of the
two following settings.

{\em Model for the phylogenetic tree on $g$ extant genera.}
For a given number $g$, we start with the species-level model 
from section \ref{sec-species-level} with the improper prior;
instead of conditioning on $n$ extant species we 
define genera as above, and condition on $g$ extant genera.
This gives a model for random phylogenetic trees which we
call e.g.
$\GENERA_g^{\theta, \mbox{\tiny fine}}$
where the superscript records the value of the
``probability of new type" parameter $\theta$ and which of
the classification schemes is used.
Now the kind of statistics for phylogenetic trees listed
in section \ref{sec-list} can in principle be studied 
within this model.

{\em Model for the phylogenetic tree of species within a genus.}
The conceptual point here is that genera are often not clades 
(some subtree forming a different genus may be absent) so that
the statistical
properties of the tree on species in a $m$-species genus will not
coincide with those for a tree on species in a $m$-species clade.
More subtly, even if a genus is a clade then the fact that 
some rule is used to select {\em which} clades are genera will 
alter statistical properties.
Our model for the tree on species in a typical extinct genus,
$\GENUS^{\theta,\mbox{\tiny fine,extinct}}$
say, is as the $n \to \infty$ limit of a randomly chosen 
genus within the $n$-species model.
Fortunately this limit interpretation constitutes a
mathematical simplification
(see ``proof strategy" in section \ref{sec-TSE}).
Similarly, our model 
$\GENUS^{\theta,\mbox{\tiny fine,extant}}$
for the tree on species in a typical extant genus
is as the $n \to \infty$ limit of a randomly chosen 
extant genus within the $n$-species model. We stress that the main focus of our results in sections \ref{sec-TSE} and \ref{sec-TSA} is on the analysis of the trees $\GENUS^{\theta,\mbox{\tiny (scheme),extinct}}$ and $\GENUS^{\theta, \mbox{\tiny (scheme),extant}}$ for the three different classification schemes.

{\em Several higher levels.}
Finally, note that it is simple to model simultaneously two or
more higher levels such as $\{ \mbox{genus, family}\}$
by using two probabilities 
$\theta_{\mathrm{family}} < \theta_{\mathrm{genus}}$.
See section \ref{sec-gen-extinct}.

\section{Mathematical results for the stochastic model}
\label{sec-analysis}
Within the stochastic model that we have defined 
we have implicitly raised 
$2 \times 2 \times 3 \times K$ 
mathematical problems; where 
$K$ is the number of interesting statistics of phylogenetic trees (cf. section \ref{sec-list}),
and where we have $2 \times 2 \times 3$ probability models
arising from different combinations of: tree on genera or tree of species within a genus; extant or extinct clades; coarse, medium or fine genera.

In the remainder of the paper we present some solutions,
emphasising those problems for which we can find reasonably
explicit analytic solutions for different genera schemes.  
Sections \ref{sec-TSE} - \ref{sec-gen-extinct} contain detailed systematic analysis in the context of extinct clades 
(which turns out to be mathematically easier).  
Section \ref{sec-TSA}  outlines one result in the extant setting.
Of course one can answer any numerical question via simulation,
and in the sequel \cite{me-macro} we identify the
most interesting features of the model for biology 
and study them via simulation where necessary.
Tables and graphs illustrating some of the formulas obtained below will
also be given in \cite{me-macro}.

\subsection{A/Z analysis}
\label{sec-AZ}
For our analysis it is more convenient to work
with ``lineage segments" than full lifelines of species.
For example, in Figure 2 the species represented by the vertical 
line
on the left has three lineage segments, determined by the two cuts
at the branchpoints where the two daughter species branched off.
Thus we can re-draw a phylogenetic tree using different lineage segments (as in Figure 9), 
where a lineage segment either ends with extinction of the species 
(or the current time), or else splits into two lineage segments,
the left one for the parent species and right one for the daughter species.
If the daughter is ``new type" this branchpoint is represented by a black circle; otherwise, a white circle.
We can now label each lineage segment as either
``type A" if some descendant species is new type,
or ``type Z" if not (A and Z are mnemonics for ``any" and ``zero").
Here the notion of any or zero ``descendants" does {\em not} include the species of the lineage segment  itself.

The advantage of this representation is that now the marks which define
the different genera schemes depend only on the
A/Z classification of the two subclades and whether the branchpoint was a new type.
Figure 9 catalogs the rules for creating marks for the three genera schemes.

\setcounter{figure}{8}
\begin{figure}
\begin{center}
\includegraphics
{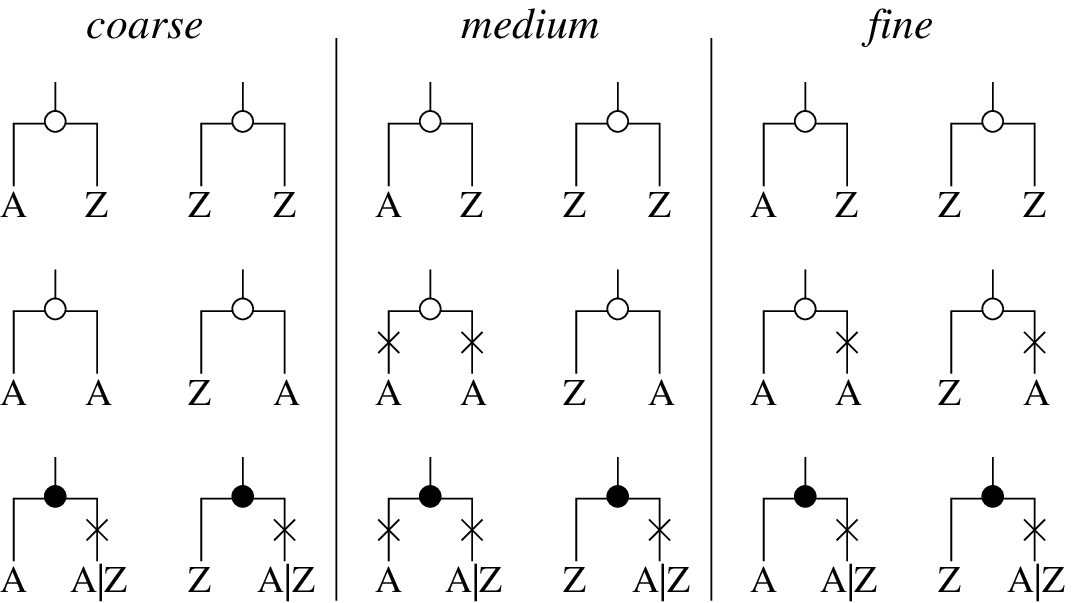}
\caption{Catalog of rules for assigning marks $\times$ in the three classification schemes. 
The figure shows a branchpoint in a phylogenetic tree redrawn as lineage segments.  Daughter lineage on right.  
Black circles
$\bullet$ 
indicate daughter is ``new type".
$Z$ or $A$ indicate zero or non-zero number of new type species in subclade.
$A|Z$ stands for "$A$ or $Z$". }
\label{ }
\end{center}
\end{figure}

\subsection{Tree of species in an extinct genus}
\label{sec-TSE}
In this section we will derive the following formulas for our probability model $\GENUS^{\theta,\mathrm{\tiny (scheme), extinct}}$.
\begin{Theorem}
\label{T2}
For the tree of species in a typical extinct genus: 
\\
(a) The mean number ($\mu$, say) of species in the genus is 
\begin{eqnarray}
&\theta^{-1} & \mbox{ ({\bf coarse})}\label{mu-coarse}\\
&\theta^{-1/2} & \mbox{ ({\bf fine})}\label{mu-fine} \\
&\theta^{-1} (3 - \theta^{1/2})^{-1}(1 + \theta^{1/2})
& \mbox{ ({\bf medium})}
 . \label{mu-medium}
\end{eqnarray}
(b) The generating function
$G(z) = Ez^\GG$
of the number $\GG$ of species in the genus is
\begin{eqnarray}
&\frac{1}{2}+\frac{1-\sqrt{(2-\theta)^2-4(1-\theta)z}}{2(1-\theta)}
&\mbox{({\bf coarse})} \label{gf-coarse}
\\
&\frac{z}{\sqrt{\theta}}
\left( \frac{1}{1 - \sqrt{\theta} + \sqrt{1-z(1-\theta)}}
- \frac{1-\theta}{1+ \sqrt{1-z(1-\theta)}} \right)
&\mbox{({\bf fine})}  \label{gf-fine}
\\
&\frac{1}{3-\sqt}
\left( 1-\sqt + 
   \frac{(1+\sqt)z-(1-\sqt)}
        {\sqrt{1-z(1-\theta)}} \right)
%
& \mbox{({\bf medium})}  \label{gf-medium}
\end{eqnarray}
(c) The distribution function 
$P(\LL \leq t)$ for the lifetime $\LL$ of the genus is
\begin{eqnarray}
& \frac{ e^{\theta t} - 1 }{ e^{\theta t} - (1-\theta ) }, & \mbox{ ({\bf coarse})}\label{LLcoarse}\\
& \frac{1}{\sqrt{\theta}}  - 
        \frac{2(1- 2 \sqt)^{-1}(1 - \sqt - \sqt e^{(2\sqt-1)t})
	- (\theta^{-1/2} - \sqt)(e^{2\sqt t} -1)}
		  {(1+\sqt) e^{2\sqt t} - (1-\sqt)}& \mbox{ ({\bf fine}) }\label{LLfine}
\end{eqnarray}
and for the medium scheme see equation (\ref{q-medium}).
\end{Theorem}

\paragraph{Proof strategy.}
The
{\em continuous time critical binary branching process} (CBP)
starts at time $0$ with one individual.
Thereafter, each species is liable to become extinct (rate $1$)
or to speciate (rate $1$).
This is of course the underlying stochastic model for species
from section \ref{sec-species-level}.
An intuitively easy result
(\cite{me111} Proposition 4)
states that, in the $n \to \infty$ limit of our $n$-species model,
the process consisting of a randomly chosen species ($\sigma$, say) and its descendants,
with time measured from the origin of species $\sigma$,
is the CBP process (call it $\TT$, say), run until extinction. At each branching point the processes continuing on either side are independent copies of the CBP. 
By now applying A/Z analysis to $\TT$,
we can assign genera to $\TT$ and then study the genus containing
$\sigma$. 
This is our proof strategy.
But be aware that
the notion of
``typical extinct genus"
does not correspond exactly to
``genus in $\TT$ containing $\sigma$". 
To make an exact correspondence, 
note (see Appendix \ref{sec-appendix}) that each genus can be identified
with a mark in the underlying $n$-species tree;
so we need to condition on $\TT$ starting with a lineage segment
which contains such a mark in the underlying tree (Figure 9).
In practice this is not difficult to do,
using the following lemma.
\begin{Lemma}
\label{LAZ}
In the CBP, the probabilities $(p_A,p_Z)$ that the initial lineage segment
are (type A, type Z) equal
\begin{equation}
 p_Z = \sfrac{1}{1+\sqt}, \quad p_A = 1-p_Z = \sfrac{\sqt}{1+\sqt}.
\label{pA}
\end{equation}
The generating function for the number $N$ of species in the CBP is
\begin{equation}
 Ez^N = 1 - \sqrt{1-z} .
\label{gf-CBP}
\end{equation}
\end{Lemma}
{\em Proof.}
Formula (\ref{gf-CBP}) is classical 
(\cite{Feller} XII.5.).
Formula (\ref{pA}) is the solution of the equation
\[ p_Z = \sfrac{1}{2} + \sfrac{1}{2} p_Z^2(1- \theta) \]
which arises as follows.
The ``$1/2$" terms are the probabilities of extinction first 
(making the lineage be type Z) or speciation.
In the latter case the only way to get type $Z$ is if both
lineages are type $Z$ and also the daughter species is not new type.

\paragraph{Derivation of formulas (a) for mean number $\mu$ of species}
In the coarse and the fine genera schemes, 
the fact that distinct genera correspond to distinct marks
implies
\begin{equation}
\mu^{-1} = P(
\mbox{typical parent-daughter edge has mark})
\label{Pmu}
\end{equation}
in the $n \to \infty$ limit.
In the coarse case, marks occur only when daughter is new type,
so this probability equals $\theta$,
giving (\ref{mu-coarse}).
The fine scheme result (\ref{mu-fine}) follows from
\begin{Lemma}
\label{LleftA}
In the $n \to \infty$ limit model, or in $\TT$, the chance that
a typical parent-daughter edge has a fine mark equals 
$\sqt$.
\end{Lemma}
{\em Proof.}
The edge does not have a fine mark if and only if the daughter
is not new type and the lineage starting with daughter is
type Z.  
So the probability in question equals
\[ 1 - (1- \theta)p_Z
= 1 - (1-\theta)/(1+ \sqt)
= 1 - (1 - \sqt) . \]

Now consider the medium scheme.
Call a lineage ``type C" if at its end the first event is speciation 
(instead of extinction) and if then both lineages are type A 
(so a type C lineage is also a type A lineage, but not necessarily
conversely).
Lemma \ref{Lmedchar} identifies medium genera with marks on lineages which are not type C. 
Then using the fact that branchpoints occur at the same rate as
daughter species, the analog of (\ref{Pmu}) 
for medium genera is
\[ \mu^{-1} = 
E
\mbox{(number of marked lineages at a typical branchpoint,} 
\mbox{ not of type C)} . \]
We now need four calculations.
\begin{eqnarray*}
P(\mbox{left lineage has medium mark})&=&
\sqt p_A \\
P(\mbox{left lineage has medium mark, is type C})&=&
\sqt \sfrac{1}{2}p_A \sqt \\
P(\mbox{right lineage has medium mark})&=&
\theta + (1-\theta)p_A^2\\
P(\mbox{right lineage has medium mark, is type C})&=&
\theta \sfrac{1}{2}p_A \sqt + (1-\theta) p_A \sfrac{1}{2}p_A \sqt 
 .
\end{eqnarray*}
Let's argue only the final one.
If the daughter is new type (chance $\theta$) we
need the daughter lineage to speciate (chance $1/2$)
and the left sublineage to be type A (chance $p_A$)
and the right sublineage to be type $A$ (chance $\sqt$ by Lemma \ref{LleftA}).
If instead the daughter is not new type (chance $1-\theta$)
then we need the original continuing lineage to be type $A$ 
(chance $p_A$) and we need the daughter's lineage to be type C 
(chance $\sfrac{1}{2}p_A \sqt$ by argument above) which implies
the original right lineage has a mark.

Using these four formulas we get
\[ \mu^{-1} = 
\sqt p_A -
\sqt \sfrac{1}{2}p_A \sqt +
(\theta + (1-\theta)p_A^2 ) - 
\left(\theta \sfrac{1}{2}p_A \sqt + (1-\theta) p_A \sfrac{1}{2}p_A \sqt  \right). \]
Using (\ref{pA}) and some manipulations we find
\[ \mu^{-1} = \frac{\theta (3 - \sqt)}{1 + \sqt} \]
leading to (\ref{mu-medium}).

\paragraph{Derivation of formulas (b) for generating functions of
number of species per genus}
First consider the coarse scheme.
Recall that for coarse genera, the MRCA of the different genera
are exactly the different ``new type" species.  
So a coarse genus consists of its ``new type" founder and its descendant species,
with the modification that any ``new type" descendant species are discarded
(and so don't have descendants).  
Because the relative chances of
a species to first (become extinct; have daughter species which is not ``new type")
are $(1; 1 - \theta)$,
it is clear that the species in a coarse genus behave as a 
Galton-Watson process whose offspring distribution $D$ is
shifted geometric$( p = 1/(2-\theta))$;
\begin{equation}
 P(D = d) = \sfrac{1}{2-\theta} \ \left(\sfrac{1-\theta}{2-\theta}\right)^d, \ d \geq 0 . \label{Dd}
\end{equation}
By classical theory (e.g. \cite{Feller} XII.5), the
probability generating function $g(z)=E(z^\GG)$ of the total 
size $\GG$ of the Galton-Watson process is determined by the
probability generating function $f_D(z) = E(z^D)$ as the unique positive 
solution of the equation $g(z)=zf_D(g(z))$.  When the 
offspring distribution is shifted geometric$(p)$ we 
have  $\ f_D(z)=p/(1-(1-p)z)$,\ and hence $\
g(z)=(1-\sqrt{1-4p(1-p)z})/2(1-p)$. 
Setting $p = 1/(2-\theta)$ gives (\ref{gf-coarse}).

We now consider fine genera. 
For a species $s$ write 
$B_s$ for the event ``neither $s$ nor any
descendant is new type".  
Writing $\sigma$ for the daughters of $s$, define
\begin{eqnarray*}
D&=& \sum_\sigma 1_{B_\sigma} \\
D^\prime &=& \sum_\sigma 1_{B_\sigma^c}\\
N_\sigma &=& \mbox{ number of species in subclade of } \sigma \mbox{ within the same genus}\\
M &=& 1 + \sum_\sigma N_\sigma 1_{B_\sigma} .
\end{eqnarray*}
Take $s$ to be a random species in the $n \to \infty$ limit of the
$n$-species model, so that the subclade of $s$ is distributed as the CBP. 
Then $s$ is the MRCA of its fine genus if and only if event $B_s^c$ occurs,
in which case the size of the genus is $M$.
So the generating function of 
``number of species per typical fine genus" is
\begin{equation}
 E(z^{M}|B_s^c)
= \frac{Ez^{M} 1_{B_s^c}}{P(B_s^c)}
= \frac{Ez^{M} - Ez^{M}1_{B_s}}{\sqrt{\theta}}  \label{zMB}
\end{equation}
because $P(B_s^c) = \sqt$ by Lemma \ref{LleftA}.

From the definition of $M$,  
\[E(z^{M})=z\sum_{d \geq 0}P(D=d)H^d(z)\] 
where $H(z)$ is the generating function of size of a clade for which event $B_s$ 
occurs, that is
\[ H(z) = E(z^M|B_s) . \]
Event $B_s$ occurs if and only if $s$ is not new type, and $D^\prime = 0$, so
\[E(z^{M}1_{B_s})=(1-\theta)z\sum_{d \geq 0}P(D=d, D^\prime=0)H^d(z).\]
Throughout the lifetime of $s$,
the chances of the next event being
\[ \mbox{($s$ goes extinct; daughter $\sigma$ and event $B_\sigma$; 
daughter $\sigma$ and event $B_\sigma^c$)} \]
are 
\[ (\sfrac{1}{2}; \sfrac{1 - \sqt}{2}; \sfrac{\sqt}{2}) \]
which easily implies
\[ P(D =d) = (1-\rho)^d\rho, \ d \geq 0; \quad \rho = 1/(2-\sqt)  \]
\[ P(D=d,D^\prime = 0) = \sfrac{1}{2} \ 
[\sfrac{1}{2}(1-\sqrt{\theta})]^d .\]
Next, because conditioning on $B_s$ is conditioning on no species to be new
type,
\[ P(M = n|B_s) = 
\frac{P(M=n) (1-\theta)^n}{P(B_s)} =
\frac{P(M=n) (1-\theta)^n}{1-\sqrt{\theta}} \] 
by Lemma \ref{LleftA} again.   
Thus, in terms of the unconditioned generating function
$G(z) = Ez^N = 1 - \sqrt{1-z}$ at 
(\ref{gf-CBP})
the conditioned generating function is
\begin{equation}
 H(z) = E(z^M|B_s) = \frac{G((1-\theta)z)}{1-\sqt} 
= \frac{1 - \sqrt{1-z(1-\theta)}}{1-\sqrt{\theta}} .\label{H-cond}
\end{equation}
We can now calculate
 \begin{eqnarray*}
  E(z^{M}) &=& z \sum_{d \geq 0} (1-\rho)^d \rho H^d(z)
 \ = \ \frac{\rho z}{1 - (1-\rho)H(z)}\\
 &=& \frac{z}{1 - \sqrt{\theta} + \sqrt{1-z(1-\theta)}} \\
 \mbox{} \quad E(z^{M}1_{B_s})
 &=& (1-\theta)z \ 
 \sum_{d\geq 0} \sfrac{1}{2} \ 
[\sfrac{1}{2}(1-\sqrt{\theta})]^d H^d(z) \\
 &=&  \ 
 \frac{(1-\theta)z \
 \sfrac{1}{2}}{1 - \sfrac{1}{2}(1-\sqrt{\theta})H(z)} 
 = \frac{(1-\theta)z}{1 + \sqrt{1-z(1-\theta)}} .
 \end{eqnarray*}
 Inserting into (\ref{zMB}) gives the desired formula (\ref{gf-fine}).

We now consider the medium scheme.
Consider the initial lineage in the CBP.
Define $M(z)$ to be the generating function
of the number of species in the CBP such that there is no medium mark
on the path between this initial lineage $s$ and the species label at
its extinction time.
Define $M_A(z), M_Z(z)$ similarly but conditioning on the
initial lineage being type $A$ or type $Z$.
So
\[
M(z) = p_A M_A(z) + p_Z M_Z(z)
\]
for $p_A, p_Z$ at (\ref{pA}); also
\begin{equation}
 M_Z(z) = H(z) =
 \frac{1 - \sqrt{1-z(1-\theta)}}{1-\sqrt{\theta}} \label{Hz}
\end{equation}
for $H(z)$ at (\ref{H-cond}) 
because each is the generating function for number of species
in the CBP conditioned on no new type species.

We first show how the desired generating function $G(z)$
of number of species in a typical extinct medium genus
is related to the quantities above.
Consider medium marks in the $n \to \infty$ limit of
the $n$-species model; recall (Lemma \ref{Lmedchar}) that we can
associate such marks with medium genera but that such genera
may be empty.
There are three possible categories of medium mark
which might be associated with a branchpoint; 
below we state their probabilities and the generating function (g.f.)
of the number of species below the mark which contribute to
the number of species in the medium genus 
associated with the mark.

Mark on parent-daughter edge where
daughter is new type: chance $ = \theta$, g.f. $= M(z)$ because  
subclade of daughter is distributed as CBP.

Daughter is not new type but parent-daughter edge has mark:
chance $= (1-\theta)p_A^2$, g.f. $ = M_A(z)$ 
because initial lineage of daughter must be type A.

Mark on continuing lineage of parent; 
chance $= p_A \sqt$ using Lemma \ref{LleftA}, 
g.f. $= M_A(z)$ because continuing lineage is type $A$.

Adding these contributions gives
\[
\bar{G}(z) = \theta M(z) 
+ \left((1-\theta)p_A^2 + p_A \sqt \right) M_A(z)
\]
whose interpretation is as the $n \to \infty$ limit
of 
$n^{-1} \sum_i z^{g_i}$ 
for the sizes $(g_i)$ of medium genera associated with marks $i$ 
in the $n$-species model.
To get the desired $G(z)$ we need to discard the null genera
and normalize to a probability distribution, so
\[ G(z) = \sfrac{
\bar{G}(z) - \bar{G}(0)}
{\bar{G}(1) - \bar{G}(0)}
 . \]
The formula for $\bar{G}$ reduces to
\[ \bar{G}(z) = \sfrac{\theta}{1+\sqt}
(M_Z(z) + 2M_A(z))
\]
and then, because $M_Z(0) = 0$,
\begin{equation}
 G(z) = 
\frac{M_Z(z) + 2(M_A(z) - M_A(0))}
{1 + 2(1-M_A(0))} . \label{G-formula}
\end{equation}
To get an equation for $M_A(z)$,
consider the initial lineage of a CBP.
In order for this to be type A, 
one of the following three possibilities must occur at the first
branchpoint; we state their (unconditional) chances and their
contribution to $M_A(z)$ if they occur.

Parent-daughter edge has no medium mark;
and either
continuing lineage is type A, daughter lineage is type Z,
or
continuing lineage is type Z, daughter lineage is type A.
Chance 
$2 (1-\theta)p_A p_Z$;
contribution to g.f. $= M_A(z)M_Z(z)$ 
because both lineages contribute.

Parent-daughter edge has a mark;
continuing lineage is type A.
Chance $p_A \sqt$; contribution to g.f. $ = 1$ because this is the
case where the genus is null.

Parent-daughter edge has a mark;
continuing lineage is type Z.
Chance $p_Z \theta$; contribution to g.f. $= M_Z(z)$
because only the continuing lineage contributes.

The unconditional probabilities 
$( 2 (1-\theta)p_A p_Z, p_A \sqt, p_Z \theta )$
of these three cases
become (after normalization to sum to $1$) the conditional
probabilities given original lineage is type A: so these conditional probabilities are 
$(
(1-\sqt),
\sfrac{1}{2} \sqt,
\sfrac{1}{2} \sqt
)$.
Thus
\[ M_A(z) = 
(1 - \sqt)M_A(z)M_Z(z) + \sfrac12\sqt + \sfrac12\sqt M_Z(z) \]
whose solution is 
\begin{eqnarray*}
M_A(z) &=& \frac{\sqt}{2} \cdot \frac{1+M_Z(z)}{1 - (1-\sqt)M_Z(z)}\\
       &=& \frac{\sqt}{2} \cdot \frac{1+M_Z(z)}{\sqrt{1-z(1-\theta )}}
\end{eqnarray*}
using (\ref{Hz}) in the denominator.
Inserting into 
(\ref{G-formula}) gives
\begin{eqnarray*}
G(z)&=&
(M_Z(z) + 2M_A(z) - \sqt)/(3 - \sqt) \\
&=& 
\frac{1}{3-\sqt}
\left( 1-\sqt + 
   \frac{(1+\sqt)z-(1-\sqt)}
        {\sqrt{1-z(1-\theta)}} \right)
\end{eqnarray*}
which is the desired expression (\ref{gf-medium}).

\paragraph{Derivation of formulas (c) for 
distribution of genus lifetime}
As before we consider the CBP. 
Write $L$ for the lifetime of the genus (in some scheme) 
containing the founding species, and write
$\type(\iota)$ for the type (A or Z) of the initial lineage
$\iota$.
Write
\begin{eqnarray*}
q_Z(t)&=& P(L \leq t, \type(\iota) = Z)
\\
q_A(t)&=& P(L \leq t, \type(\iota) = A)
 . 
\end{eqnarray*}
We shall argue that these distribution functions satisfy
the differential equations
\begin{eqnarray}
 q^{\prime}_Z + 2 q_Z &=& (1-\Th) q_Z^2 + 1, \label{qZ-de}
\\
 (q^{\prime}_Z+q^{\prime}_A) + 2(q_Z+q_A) &=& R(q_A,q_Z,p_A) +1 
\label{qA-de}
\end{eqnarray}
with initial conditions 
\[ q_Z(0) = q_A(0) = 0, \]
where $p_A = \sqt / (1+\sqt)$ 
(\ref{pA})
and where $R$ is a function 
depending on genera scheme, given in 
(\ref{eq.Rc},\ref{eq.Rf}), 
for the coarse and fine scheme below.
These equations are derived using the
{\em backwards equations} method for branching processes
(\cite{Feller}, XVII.8).
That is, in initial time $dt$ we have 
(to first order in $dt$)\\
\hspace*{0.3in}
chance $1 \cdot dt$ of extinction\\
\hspace*{0.3in}
chance $1 \cdot dt$ of a branchpoint\\
\hspace*{0.3in}
chance $1 - 2 \cdot dt$ of neither.

\noindent
To derive (\ref{qZ-de})
note that the ``extinction" possibility implies the lineage is type Z.
So,
\begin{equation}
q_Z(t+dt) = 1 \cdot dt 
+ \Xi_t \ 1 \cdot dt
+ (1 - 2 \cdot dt)q_Z(t)
\label{qZ1}
\end{equation}
where 
\[ \Xi_t = P(L \leq t, \type(\iota) = Z| 
\mbox{branchpoint at time } 0+) . \]
This rearranges to
\begin{equation}
q^\prime_Z(t) + 2 q_Z(t) = 1 + \Xi_t .
\end{equation}
In order to have 
$L \leq t$ and $\type(\iota) = Z$
after a branchpoint at time $0+$, 
we need the daughter to not be new type, and we need
both subsequent lineages to be type Z, so
$\Xi_t = (1-\theta) q_Z^2(t)$, 
giving (\ref{qZ-de}).

To derive (\ref{qA-de}), repeat the argument for (\ref{qZ1}) 
to get
\[
q^\prime_Z(t) + q^\prime_A(t) 
+ 2(q_Z(t) + q_A(t))
= 1 + R_t 
\]
\[ R_t = P(L \leq t | 
\mbox{branchpoint at time } 0+) . \]
We now consider each scheme in turn,
to get formulas for $R_t$.
In each case we condition on a branchpoint at time $0+$, 
and write $L_l, L_r$ 
for the ``lifetime of genus"
quantity $L$ applied to the lineages $\iota_l, \iota_r$ after the branchpoint.

\noindent
{\bf For coarse genera}, 
in order to have $L \leq t$ we need \\
\hspace*{0.3in} 
either: daughter is new type, and
$L_l \leq t$;\\
\hspace*{0.3in}
or: daughter is not new type, and $L_l \leq t$,  and $L_r \leq t$.

\noindent 
So,
\begin{equation}
R_t = \theta (q_Z(t) + q_A(t)) + (1-\theta)(q_A(t)+q_Z(t))^2
:= 
R_{coarse}(q_A(t),q_Z(t)) .
\label{eq.Rc}
\end{equation}

\noindent
{\bf For fine genera}, 
in order to have $L \leq t$ we need \\
\hspace*{0.3in}
either: daughter is new type, and $L_l \leq t$;\\
\hspace*{0.3in}
or: daughter is not new type, and $L_l \leq t$, and
$\type(\iota_r) = A$;\\
\hspace*{0.3in}
or: daughter is not new type, and $L_l \leq t, \type(\iota_r) = Z$,  and $L_r \leq t$
 .

\noindent
So,
\begin{equation}
 R_t = 
\theta (q_A(t) + q_Z(t))
+ (1 - \theta) (q_A(t) + q_Z(t)) (p_A + q_Z(t))
:= R_{fine}(q_A(t),q_Z(t),p_A) .
\label{eq.Rf}
\end{equation}

\noindent The case of medium genera is more complicated
and will be treated below separately.


Solving these equations we get the following results:
\begin{eqnarray}
q_Z(t) &=& \frac{ e^{2\sqt t} - 1 }
           { (1+\sqt) e^{2\sqt t} - (1-\sqt) }%
\label{qZ}
\\
  q_Z(t) + q_A(t) \,\Big|_{coarse} &=& \frac{ e^{\theta t} - 1 }{ e^{\theta t} - (1-\Th) } 
\label{qAZcoarse}
\\
q_A(t)+q_Z(t) \,\Big|_{fine} &=&
    1 - \frac{2\sqt}{1-2\sqt} \cdot
        \frac{(1-\sqt) - \sqt e^{(2\sqt-1)t}}
		  {(1+\sqt) e^{2\sqt t} - (1-\sqt)}.
\label{qAZfine}
\end{eqnarray}

Note that the branching process of a Z lineage is a birth-death process 
with birth rate $1-\sqrt{\theta}$ and death rate $1/p_z=1+\sqrt{\theta}$ so (\ref{qZ}) also follows from the standard result on the lifetime distribution of a birth-death process (\cite{Feller} XVII.10.ex.12). Also, the branching process of the coarse genera is a birth-death process with birth rate $1-\theta$ and death rate $1$, and (\ref{qAZcoarse}) follows from the lifetime distribution of such birth-death process. 

We now need to translate these 
formulas 
$P(L \leq t) = q_Z(t) + q_A(t)$
for the distribution function of $L$
(the size of genus in a CBP containing the founder species $\sigma$)
into
formulas for the distribution function of $\LL$
(the size of a typical extinct genus), the relation being that 
$\LL$ has the conditional distribution of $L$ given that 
$\sigma$ 
(regarded as a species sampled from an $n \to \infty$ limit clade)
is founder of a genus in that clade.
For coarse genera, we are just 
conditioning on $\sigma$ being a new type species, which has no effect on 
the distribution of the CBP, so formula 
(\ref{qAZcoarse}) immediately becomes formula (\ref{LLcoarse}) for $\LL$.

For fine genera, the marking rule implies
\begin{quote}
$\LL$ has the conditional distribution of $L$ given that 
either $\sigma$ or some descendant of $\sigma$ is new type.
\end{quote}
There is chance $\theta$ for $\sigma$ to be new type,
and chance $(1- \theta )p_A$ that $\sigma$ is not new type but some
descendant is new type. 
So 
\begin{eqnarray*}
P(\LL \leq t)
&=&
\frac{\theta}
{\theta + (1-\theta)p_A}
(q_A(t) + q_Z(t))
+ \frac{(1-\theta) p_A}
{\theta + (1-\theta)p_A}
\frac{q_A(t)}{p_A}
\\ &=&
\frac{q_A(t) + q_Z(t) - (1-\theta)q_Z(t)}{\sqt}
\end{eqnarray*}
using the fact 
${\theta + (1-\theta)p_A}
= \sqt$.
Now (\ref{qZ},\ref{qAZfine}) give formula (\ref{LLfine}).


\

\noindent
{\bf For medium genera.}
We identify each genus with the oldest species in the genus
(for the fine and coarse schemes this is essentially the same as identifying
a genus with a mark on a tree, but not for the medium scheme),
the birth time of this oldest species giving the ``starting point" 
from which we measure genus lifetime.

Consider a species $a$ that originates from its parent $b$ at the 
branchpoint $\beta$.
For $a$ to be the oldest species in its genus
there are three alternatives:\\
(1) 
there is a medium mark on the lineage of $a$;
in this case we say that the right subtree below $\beta$ has type $B$;
\\
(2)
there is a medium mark on the parent-daughter edge 
$b$ - $a$,
but no marks on the lineage of $a$;
\\
(3)
there are no marks on the lineage of $a$ and 
no medium mark on the parent-daughter edge
$b$ - $a$,
but $a$ is still the oldest in its genus.

It's clear that in cases (1,2) no species older than
$a$ can be in the same genus
(because by definition the two species are in the same genus
only if the path in the tree between the corresponding leaves 
contains no marked edge).
Consider in detail how case (3) can arise.

Because $a$ and $b$ are in different genera, but there is no medium mark
on the path from $a$-leaf to branchpoint $\beta$, there is necessarily
a mark between $\beta$ and $b$-leaf.   In this case we say that the left tree below $\beta$ 
has type $B$.
Note that this means that $a$ necessarily has type $Z$; otherwise $\beta$ would
be a branchpoint of type $A+A$ and both edges below it would have medium marks.
If the branchpoint directly above $\beta$ is the starting point of another 
daughter of $b$ (this has probability $1/2$), 
this daughter must be either new type or type $A$
(probability $(1-\theta)p_A + \theta = \sqt$); otherwise it would be in the same
genus as $a$ and $a$ would not be the oldest in its genus.
If the branchpoint $\beta'$ directly above $\beta$ 
is the starting point of $b$ itself (probability $1/2$),
then (denoting by $b'$ the parent of $b$) 
for $a$ and $b'$ to be in different genera,
either the parent-daughter edge $b'$ - $b$ must have a medium mark,
or the segment of the lineage of $b'$ between $\beta'$
and $b'$-leaf must have a medium mark. 
But in the latter case, both subtrees below $\beta'$ are of type $A$,
so there is a mark on parent-daughter edge $b'$ - $b$ anyway
(probability $\theta + (1-\theta)p_A = \sqt$).

The probability for the initial lineage to be of type $B$ 
satisfies the equation
\[ p_B = \sfrac{1}{2} p_A \sqt + \sfrac{1}{2} (1-\sqt) p_B, \]
so the first alternative has probability
\[ p_B = \frac{\theta}{(1+\sqt)^2}, \]
the second alternative has probability
\[ \theta (1-p_B) + (1-\theta) (p_A-p_B) p_A =
   \frac{ \theta(\sqt+2) }{ (1+\sqt)^2 },
\]
and the third alternative has probability
\[ p_Z p_B (\sfrac{1}{2} \sqt + \sfrac{1}{2} \sqt -\theta ) = \frac{\theta(\sqt-\theta)}{(1+\sqt)^2}. \]
Summing these three contributions, the
probability that a randomly chosen species 
is the oldest in its genus equals
$ \frac{\theta(3-\sqt)}{1+\sqt}$,
which agrees with (\ref{mu-medium}).

Now let's return to the lifetime distribution.
As before we write $L$ for the lifetime of the genus
containing the founding species, and write
$\type(\iota)$ for the type of the initial lineage $\iota$.
We'll need three types: Z, B (which is subset of A)
and $\bA$ (which stands for ``$A$ but not $B$").
Write
\begin{eqnarray*}
q_Z(t)&=& P(L \leq t, \type(\iota) = Z)
\\
q_\bA(t)&=& P(L \leq t, \type(\iota) = \bA)
\\
q_B(t)&=& P(L \leq t, \type(\iota) = B)
. 
\end{eqnarray*}
$q_Z(t)$ was calculated earlier.
$q_B(t)$ is defined by
\[ q_B'(t) + 2 q_B(t) = q_B(t) + \theta q_\bA(t) + (1-\theta)p_A q_\bA(t),
 \qquad q_B(0)=0,
\]
$q_\bA$ is defined by
\[ q_\bA'(t) + 2 q_\bA(t) = \theta q_Z(t) + 2(1-\theta) q_Z(t)q_\bA(t)
   + (1-\theta) q_Z(t) q_\bB(t), 
  \qquad q_\bA(0)=0,
\]
where $q_\bB(t)$ is defined by
\[ q_\bB'(t) + 2 q_\bB(t) = (1-\theta) q_\bB(t) q_Z(t) + \sqt p_A,
  \qquad q_\bB(0) = 1/2 p_A \sqt.
\]
Here $q_\bB$ is the probability that the lineage $\iota$ has type $B$
and the oldest leaf, reachable from the initial point of the CBP
(that is such that there is no medium marks on the path to this leaf)
is not older than $t$.
Because it's possible that the first branchpoint in the CBP
is of type $A+A$, and no leaves are reachable from the origin,
we need a non-zero initial condition for $q_\bB$.

Solving the differential equations above, we find
\[ q_\bB(t) = 
   \frac{ \theta(e^{2t\sqt}-1) + \theta^{3/2} e^{t\sqt - t}}
        { (1+\sqt)^2 e^{2t\sqt} + \theta -1 },
\]
\[ q_\bA(t) = \frac
{ \psi(\theta,t) }
{ ( (1+\sqt) e^{2t\sqt} + \sqt -1 )^2 (1+\sqt) },
\]
\begin{eqnarray*}
\psi(\theta,t) &=& 
\left (\theta+\sqrt {\theta}\right ){e^{4 t\sqrt {\theta}}}+\theta 
\left (2 \theta-2 t\sqrt {\theta}-4 t-\sqrt {\theta}+2 t\theta-1
\right ){e^{2 t\sqrt {\theta}}}
\\
&& {}
+{\theta}^{3/2}-\sqrt {\theta}-\theta
 \left (\theta {e^{2 t\sqrt {\theta}}}+{e^{2 t\sqrt {\theta}}}
\sqrt {\theta}+\theta-\sqrt {\theta}\right ){e^{t\left (\sqrt {\theta}
-1\right )}}
\end{eqnarray*}
and
\[ q_B(t) = \sqt \int_0^t q_\bA(u) e^{u-t} \, du. \]

Finally, the overall genus lifetime distribution function is
a weighted sum of distributions above, giving
\begin{equation} \label{q-medium} 
   P(\LL \leq t)  = 
   \frac{ 1+\sqt }{\theta(3-\sqt)} \Big(q_B(t) +
   \theta q_Z(t) + \sqt q_\bA(t)
   + \sqt (1-\theta) q_\bB(t) q_Z(t) \Big).
\end{equation}

\subsection{Tree of extinct genera}
\label{sec-gen-extinct}
We now consider aspects of the trees on genera, 
illustrated in Figures 7 and 8.
For the first result, each genus has some number (maybe zero)
of ``direct offspring" genera.
For instance, in Figure 7 (fine genera) genus $b$ has two offspring, the
first genus $c$ and the genus $k$.
For Figure 7 the numbers of fine genera with 
$(0;1;2)$ offspring genera are $(4; 18; 3)$.
Let us write ``offspring tree" for the tree recording this ``direct offspring" relationship
between genera.
So the offspring tree carries less information than the complete tree 
(e.g. the lifetime of a genus is not included)
but more information than the induced cladogram 
(e.g. it includes the identity of MRCA genera).
\begin{Proposition}
\label{P4}
The offspring tree on descendant genera of a typical extinct genus is 
distributed as a Galton-Watson branching process, 
whose offspring distribution $\xi$ is as follows.\\
(a) {\bf Coarse genera:}
\begin{eqnarray*}
P(\xi=0)&=&p_Z\\
P(\xi=i)&=&\frac{\theta P(\xi=i-1)+(1-\theta)\mathop{\sum}_{j=1}^{i-1}P(\xi=j)P(\xi=i-j)}{2\sqt}, \ i\geq 1
\end{eqnarray*}
where $p_Z=1/(1+\sqt)$.
\\
(b) {\bf Fine genera:}
\begin{eqnarray*}
P(\xi = 0) &=& p_A \\
P(\xi = i) &=& (1-p_A)^2p_A^{i-1}, \ i \geq 1 
\end{eqnarray*}
where $p_A = \sqt /(1 + \sqt)$.  
\end{Proposition}
For medium genera there is a more complicated result (which we omit) involving  a three-type Galton-Watson process.
Note $E \xi = 1$, so that (as expected) the ``critical" property
of the species-level model is preserved at the genus level.
Note also that in Figure 7, where $p_A = 1/6$, the data on offspring 
frequency matches well the distribution (b) of $\xi$, even though Figure 7
refers to the extant setting.

{\em Proof.}
(a) For coarse genera, each genus is founded by a new type species,
so clearly the offspring tree we seek is the Galton-Watson process
with offspring distribution $\xi$ described as
follows:
\begin{quote}
start CBP with a species $\sigma$ which is not new type, but disallow descendants
of any new type species; let $\xi$ be the number of new type species.
\end{quote}
Because $\sigma$ may 
(become extinct; have new type daughter; have not new type daughter)
with chances
$(1/2; \theta /2; (1-\theta)/2)$,
the generating function
$\Phi(z) := E z^\xi$
satisfies the equation
\[ \Phi = \sfrac{1}{2}
\left( 1 + \theta z \Phi + (1-\theta) \Phi^2 \right) \]
whose solution is 
\begin{equation}
\Phi(z)=E z^\xi = \frac{
1 - \frac{1}{2}\theta z - 
\sqrt{\frac{1}{4}\theta^2z^2 - \theta z + \theta}
}{1 - \theta} .
\label{gf-GW}
\end{equation}
One can deduce the recursive formula stated in (a).

(b) Consider a species $\sigma$ as the founder of CBP
and as a sampled species from a large clade conditioned on
the edge 
$(\mbox{parent}(\sigma),\sigma)$
having a fine mark.
Consider the fine genus $g$ founded by $\sigma$.
The number of 
offspring genera is exactly the number $\xi$ of daughter species
$\sigma_i$ of $\sigma$
such that the edge 
$(\sigma,\sigma_i)$ 
has a fine mark. 
It easily follows that the
offspring tree under consideration in Proposition \ref{P4} is a Galton-Watson
process with some offspring distribution $\xi$.
Recall the $A/Z$ analysis from section \ref{sec-AZ}.
If the initial lineage of $\sigma$ is type Z then $\xi = 0$.
If it is type A (probability $q$, say) then at each marked edge 
$(\sigma,\sigma_i)$ 
there is some probability ($r$, say)
that the continuing lineage of $\sigma$ is type A.
So the distribution of $\xi$ has the form
\begin{eqnarray*}
P(\xi = 0) &=& 1-q \\
P(\xi = i) &=&
qr^{i-1}(1-r), \ i \geq 1 .
\end{eqnarray*}
To calculate $r$, note that 
conditioning on a parent-daughter edge having a fine mark 
(which forces the lineage above the split to be type A)
does not affect probabilities for the type of the continuing parental-species 
lineage, so $r = p_A$ in 
(\ref{pA}).
To calculate $q$, in the setting of the founder $\sigma$ of CBP,
\[ q = P(
\mbox{lineage is type A}
|
\mbox{lineage is type A, or $\sigma$ is new type}) . \]
The conditioning event has chance $\sqt$ by Lemma \ref{LleftA}.
So
\[ q = p_A/\sqt = p_Z = 1-p_A  \]
giving the distribution in (b).
\

\paragraph{Interpretation}
There are several ways to interpret Proposition \ref{P4}
as a statement about 
``typical trees on extinct genera".
First, we could consider the tree on all genera in a large clade; 
given that a subtree has $g$ genera, this subtree
(that is, its tree of offspring) 
is the Galton-Watson process in Proposition \ref{P4}
conditioned on having exactly $g$ genera.

Another interpretation uses the genus/family 
model mentioned at the end of
section \ref{sec-pmhot}.
Set 
$\theta_{\mathrm{family}} < \theta =  \theta_{\mathrm{genus}}$
and suppose that each species has chance 
$\theta$ to be new genus type or new family type,
and chance 
$\theta_{\mathrm{family}} $ 
to be new family type.
Now we can consider
``the tree on genera in a typical family" 
in the way analogous to 
``the tree on species in a typical genus" 
previously studied.

\begin{Proposition}
\label{P5}
(a) In the {\bf coarse} scheme, 
the offspring tree on genera in a typical extinct family 
is distributed as the Galton-Watson branching process whose
offspring distribution $\xi^\prime$ 
has generating function
\begin{equation}
 \Phi^\prime(z) = \Phi(z + \theta^\prime (1-z)) \label{gf-GW2}
\end{equation}
where
$ \theta^\prime = 
\theta_{\mathrm{family}}  / \theta  
$
and where
$\Phi(z)$ is the generating function 
(\ref{gf-GW}).\\
(b) In the {\bf fine} scheme, the offspring tree on genera in a
typical extinct family is a Galton-Watson branching process with
offspring distributions $\eta_0, \eta$ as follows.  
After the first generation the offspring distribution $\eta$ is determined by the relation 
\begin{equation}
\frac{P(\eta = i)}{P(\eta = 0)} = \theta^{-1} \left( \frac{\sqrt{\theta} - \sqrt{\theta_{\mathrm{family}}}}{1+\sqrt{\theta}} \right)^i, \quad i \geq 1 . 
\label{eet}
\end{equation}
In the first generation, 
\begin{equation}
P(\eta_0 = i) = \sfrac{ \sqrt{\theta_{\mathrm{family}}}}{1 + \sqrt{\theta_{\mathrm{family}}}} 
P(\eta = i) + \sfrac{ 1}{1 + \sqrt{\theta_{\mathrm{family}}}}  P(\zeta = i) 
\end{equation}
where 
\begin{equation}
 P(\zeta = i) \propto \left(\sqrt{\frac{\theta}{\theta_{\mathrm{family}}}} - 1\right)^i  \ 
\sum_{j \geq i+1} {j \choose i}  \left( \frac{\sqrt{\theta_{\mathrm{family}}}}{1+\sqrt{\theta}} \right)^j, \  i \geq 0 . \label{zeta}
\end{equation}
\end{Proposition}
{\em Proof.}
(a) The process of all descendant species of a typical
``new family type" species $\sigma$
is just CBP.
So as in Proposition \ref{P4}(a), 
the offspring tree of genera 
(where a genus may or may not be in a new family)
is just 
the Galton-Watson process with offspring distribution $\xi$
at (\ref{gf-GW}).
We want the subprocess containing only genera in the same family as $\sigma$.
Because a new genus type has chance $\theta^\prime$ to be a new family type,
the subprocess is just 
the Galton-Watson process with offspring distribution $\xi^\prime$
described by
\begin{quote}
the conditional distribution of $\xi^\prime$ given $\xi$
is Binomial$(\xi, 1 - \theta^\prime)$ 
\end{quote}
and (\ref{gf-GW2}) follows.

(b) In the fine scheme, a species $\sigma$ founds a new genus (resp. family) if the 
parent-daughter edge $(\sigma^\prime, \sigma)$ has a genus (resp. family) mark, which by Lemma \ref{LleftA} has probability $\sqrt{\theta}$ (resp. $ \sqrt{\theta_{\mathrm{family}}}$).  
Here ``mark" means {\em fine} mark.  
So in particular,

given $(\sigma^\prime, \sigma)$ has a genus mark, the chance it has a family mark 
equals $\sqrt{\theta_{\mathrm{family}}/\theta} = \rho$, say.
\\
Consider first the case where  $(\sigma^\prime, \sigma)$ has a genus mark but no family mark. 
In this case all descendant genera are in the same family, and the number $\eta$ of offspring genera 
has distribution
\[ P(\eta = i) \propto P(\xi = i) (1 - \rho)^i \]
for $\xi$ as in Proposition \ref{P4}(b). 
Now consider the case where  $(\sigma^\prime, \sigma)$ has a family mark, so that the genus of
$\sigma$ is the founder genus in a new family.  A daughter genus $\sigma_*$ for which the edge 
$(\sigma,\sigma_*)$ has a family mark will be in a different family.  Thus the distribution 
$\eta_0$ for number of offspring genera within a family for the founding genus can be described as:

conditional on $(\sigma^\prime, \sigma)$ having a family mark, $\eta_0$ is the number of offspring genera without a family mark.

\noindent
To get an expression for the distribution of $\eta$, use Proposition \ref{P4}(b) to get, for $i \geq 1$,
\[ \frac{P(\eta = i)}{P(\eta = 0)} = \left( \sfrac{1 - p_A}{p_A}\right)^2 \ (p_A(1- \rho))^i
= \theta^{-1} \left( \frac{\sqrt{\theta} - \sqrt{\theta_{\mathrm{family}}}}{1+\sqrt{\theta}} \right)^i  \] 
which is (\ref{eet}). 

So consider the case where  $(\sigma^\prime, \sigma)$ has a family mark. 
This splits into two sub-cases:\\
(i) some descendant of $\sigma$ is new family type;\\
(ii) $\sigma$ itself, but no descendant, is new family type.\\
By considering a typical species $\sigma$ and using Lemma \ref{LAZ}, the relative chances 
of (i) and (ii) are $\frac{\sqrt{\theta_{\mathrm{family}}}}{1 + \sqrt{\theta_{\mathrm{family}}}}$ 
and $\frac{1}{1 + \sqrt{\theta_{\mathrm{family}}}} \times \theta_{\mathrm{family}}$, so the actual chances are $\frac{1}{1 + \sqrt{\theta_{\mathrm{family}}}} $ and $\frac{\sqrt{\theta_{\mathrm{family}}}}{1 + \sqrt{\theta_{\mathrm{family}}}}$.
We are interested in the number $\eta_0$ of offspring genera in the same family, which in sub-case (ii) is the same as $\eta$ above.  In sub-case (i) the number $\zeta$ of same-family offspring genera can be written as 
\[ P(\zeta = i) = P(\xi^\prime = i|\xi^{\prime \prime} \geq 1) \]
where $(\xi^\prime,\xi^{\prime \prime})$ are the number of (not new family, new family) offspring genera of a species $\sigma$ founding a genus.   
Now $\xi^\prime+\xi^{\prime \prime}$ has the distribution of $\xi$ in Proposition \ref{P4}(b), and conditionally on $\xi^\prime+\xi^{\prime \prime}$ each genus has probability $\rho$ to represent a new family, so
\begin{eqnarray*}
P(\zeta = i) &\propto&
\sum_{j\geq i+1} P(\xi = j) {j \choose i} (1-\rho)^i \rho^{j-i} \\
&\propto& (\sfrac{1}{\rho} -1)^i \sum_{j\geq i+1} (p_A \rho)^j  {j \choose i} 
\end{eqnarray*}
which leads to (\ref{zeta}).

\subsection{Extant clades}
\label{sec-TSA}
In principle the previous calculations could be repeated in the (more interesting, perhaps) setting of extant clades.  However, this is more complicated because in the context of Theorem \ref{T2} it is now natural to condition of the number $n$ of extant species.  Similarly, in the context of Proposition \ref{P4} it is now natural to condition of the number $g$ of extant genera.  These extra parameters must make explicit formulas more complicated and we have not attempted systematic analysis.  Let us just give one result avoiding such conditioning which can be proved by a clever trick.
\begin{Proposition}
\label{P6}
In the {\bf coarse} scheme, 
the number $\tilde{N}$ of extant species in 
the genus of a typical extant species 
has Geometric$(\theta)$ distribution
\begin{equation}
P(\tilde{N} = n ) = \theta (1-\theta)^{n-1}, \quad n \geq 1 .
\label{Ntilde}
\end{equation}
Equivalently,  the number $N$ of extant species in 
a typical extant genus 
has the inverse-size-biased distribution
\begin{equation}
P(N=n) = \frac{(1-\theta)^n}{c_\theta n}, \quad n \geq 1 
\label{Nnn}
\end{equation}
where 
$c_\theta = \sum_{n \geq 1} (1-\theta)^n/n = - \log \theta$. 
So
\[
E N = \sfrac{\theta - 1}{\theta \log \theta} .
\]
\end{Proposition}
{\em Proof.}
Consider the CBP where the initial species $\sigma$ is new type.
Let $X_t$ be the number of species alive at time $t > 0$ which are in the same coarse
genus as $\sigma$.
Then
\begin{equation}
P(N = \cdot) \propto \int_0^\infty P(X_t = \cdot) \ dt
\label{PNpt}
\end{equation}
because in the underlying infinitely-large clade, new type species arose at constant rate in the past.
Now $(X_t)$ is the birth-and-death (continuous-time Markov) process on states $0,1,2,\ldots$, 
started at state $1$, with transition rates
\[ q(x,x+1) = (1-\theta)x; \quad \quad 
q(x,x-1) = x. \]
By Markov chain theory the ``mean occupation time" in (\ref{PNpt}) is proportional
to the stationary distribution $\pi(\cdot)$ of the process 
$(X_t)$
(after we insert some arbitrary transition rate $0 \to 1$).
But the stationary distribution satisfies
\[ \pi(x+1)/\pi(x) = q(x,x+1)/q(x+1,x)
= (1-\theta)x/(x+1)
\]
whose solution is (\ref{Nnn}).

\paragraph{Acknowledgement}  
We thank a referee for helpful expositiory suggestions.


\def\cprime{$'$} \def\polhk#1{\setbox0=\hbox{#1}{\ooalign{\hidewidth
  \lower1.5ex\hbox{`}\hidewidth\crcr\unhbox0}}} \def\cprime{$'$}
  \def\cprime{$'$} \def\cprime{$'$}
  \def\polhk#1{\setbox0=\hbox{#1}{\ooalign{\hidewidth
  \lower1.5ex\hbox{`}\hidewidth\crcr\unhbox0}}} \def\cprime{$'$}
  \def\cprime{$'$} \def\polhk#1{\setbox0=\hbox{#1}{\ooalign{\hidewidth
  \lower1.5ex\hbox{`}\hidewidth\crcr\unhbox0}}} \def\cprime{$'$}
  \def\cprime{$'$} \def\cydot{\leavevmode\raise.4ex\hbox{.}} \def\cprime{$'$}
  \def\cprime{$'$} \def\cprime{$'$} \def\cprime{$'$} \def\cprime{$'$}
  \def\cprime{$'$} \def\cprime{$'$} \def\cprime{$'$} \def\cprime{$'$}
  \def\cprime{$'$}

 \end{document}